\documentclass[3p,10pt]{elsarticle}
\usepackage{amsmath}

\usepackage[caption=false]{subfig}
\usepackage{graphicx}
\usepackage{epsfig}
\usepackage{amssymb}
\usepackage{MnSymbol}
\usepackage{multirow}
\usepackage{bigdelim}
\usepackage{slashed}
\usepackage{enumerate}
\usepackage{epstopdf}
\usepackage{epsfig}
\usepackage{enumerate}
\usepackage[colorlinks]{hyperref}
\usepackage{comment}
\usepackage[usenames, dvipsnames]{color}
\usepackage{hyperref}
\usepackage[T1]{fontenc}
\usepackage{pstricks}
\usepackage{hyperref}
\makeatletter
\def\ps@pprintTitle{%
	\let\@oddhead\@empty
	\let\@evenhead\@empty
	\def\@oddfoot{\centerline{\thepage}}%
	\let\@evenfoot\@oddfoot}
\makeatother

\biboptions{sort&compress}
\newcommand{\bea}{\begin{eqnarray}}
\newcommand{\eea}{\end{eqnarray}}
\newcommand{\bean}{\begin{eqnarray*}}
\newcommand{\eean}{\end{eqnarray*}}
\newcommand{\nn}{\nonumber \\}

\def\njet{{{\sc NJet}}}

\def\f{\tilde{f}}

\def\Feynarts{{{\sc FeynArts}}}

\def\mathematica{{{\sc mathematica}}}
\def\sam{{{\sc S@M}}}

\def\Label#1{\label{#1}%
  \smash{\hbox to0pt{\raise1ex\hbox{\tiny[#1]}\hss}}}

\allowdisplaybreaks

\newcommand{\unipd}{Dipartimento di Fisica ed Astronomia, Universit\`a di Padova, Via Marzolo 8, 35131 Padova, Italy}
\newcommand{\pdinfn}{INFN, Sezione di Padova, Via Marzolo 8, 35131 Padova, Italy}
                     
\journal{Physics Letters B}
\begin{document}
\begin{frontmatter}
\title{BCJ Identities and $d$-Dimensional Generalized Unitarity
	}
\address[pd]{\unipd}
\address[infn]{\pdinfn}

\author[pd,infn]{Amedeo Primo}
\ead{amedeo.primo@pd.infn.it}
\author[pd,infn]{William J. Torres Bobadilla}
\ead{william.torres@pd.infn.it}

\date{\today}

\begin{abstract}
	We present a set of relations between one-loop integral coefficients for dimensionally regulated QCD amplitudes. Within dimensional regularization, the combined use of color-kinematics duality and integrand reduction yields the existence of relations between the integrand residues of partial amplitudes with different orderings of the external particles. These relations can be established for the cut-constructible contributions as well for the ones responsible for rational terms.\\
	Starting from the general parametrization of one-loop residues and applying Laurent expansion in order to extract the coefficients of the amplitude decomposition in terms of master integrals, we show that the full set of relations can be obtained by considering BCJ identities between $d$-dimensional tree-levels.\\
	 We provide explicit examples for multi-gluon scattering amplitudes at one-loop.

\end{abstract}
\end{frontmatter}
\input{feynarts.sty}

\section{Introduction}
Tree-level amplitudes in gauge theories are known to satisfy color-kinematics (C/K) duality, \textit{i.e.} they admit an expansion in terms of Feynman diagrams where the kinematic parts of the numerators satisfy the same antisymmetry and Lie-Algebra identities as their corresponding color factors.\\ This property was first observed  by Bern, Carrasco and Johansson for pure gauge amplitudes in ~\cite{Bern:2008qj,Bern:2010ue} and later extended to both massless and massive QCD, ~\cite{Johansson:2014zca,Naculich:2014naa,Johansson:2015oia}.\\
One of most striking implications of C/K duality is the existence of 
relations between color-ordered tree-level amplitudes~\cite{Bern:2008qj}, which, together with $U(1)$ symmetry and Kleiss-Kuijf relations~\cite{Kleiss:1988ne}, can be used to further reduce the number of independent partial amplitudes to be considered in tree-level calculations. \\
In \cite{Mastrolia:2015maa}, by adopting the Four-Dimensional-Formulation (FDF)~\cite{Fazio:2014xea} variant of the Four-Dimensional-Helicity (FDH)~\cite{Bern:1991aq,Bern:1995db,Bern:2002zk} regularization scheme, we studied C/K-duality for tree-level amplitudes in $d$-dimensions and we derived a set of BCJ identities, for four- and five-point amplitudes, which take into account the explicit dependence on the regulating parameter, together with a general strategy for the determination of analogous relations between higher-multiplicity amplitudes.\\

The recent development of on-shell~\cite{Cachazo:2004kj,Britto:2004ap} and generalized unitarity techniques~\cite{Bern:1994zx} for quadruple-\cite{Britto:2004nc,Badger:2008cm}, triple-\cite{Badger:2008cm,Mastrolia:2006ki,Forde:2007mi}, double-\cite{Britto:2006sj,Mastrolia:2009dr} and single-\cite{Kilgore:2007qr,Britto:2009wz,Britto:2010um} cut allowed tremendous simplifications in one-loop calculations, where the knowledge of tree-level amplitudes can be exploited in order to determine the coefficients of the known basis of integrals in which any amplitude can be decomposed, ~\cite{Passarino:1978jh,vanOldenborgh:1989wn}.\\
In the framework of four-dimensional generalized unitarity, BCJ identities for tree-level amplitudes were used in \cite{BjerrumBohr:2010zs} to derive relations between coefficients of one-loop amplitudes in $\mathcal{N}=4$ super Yang-Mills theory and, more recently, in \cite{Chester:2016ojq} these relations have been extended to integral coefficients for the cut-constructible part of one-loop QCD amplitudes, showing that tree-level C/K-duality can significantly decrease the number of independent coefficients needed in one-loop computations. \\

In this paper, by making use of BCJ identities for dimensionally regulated trees, we provide a set of coefficient relations for one-loop QCD amplitudes which include the contributions from rational terms.\\
The paper is organized as follows: in Section~\ref{sec:FDFduality} we recall the main results regarding BCJ identities for tree-level amplitudes in $d$-dimensions, obtained by using the FDF scheme. In Section ~\ref{sec:coef-relations} we review the decomposition of one-loop amplitudes via integrand reduction~\cite{Ossola:2006us,Ellis:2007br,Mastrolia:2011pr,Badger:2012dp,Zhang:2012ce,Mastrolia:2012an,Mastrolia:2013kca} and we apply $d$-dimensional BCJ identities between four-point amplitudes in order to establish general relations between the coefficients appearing in the decomposition. In Section  ~\ref{sec:example}, we verify the coefficient identities in a few concrete examples, by showing relations between the analytic expression of the coefficients for scalar-loop contributions to multi-gluon amplitudes, up to six-points. \\
Finally, in \ref{apd} we extend the results of Section~\ref{sec:coef-relations} by providing the set of coefficient relations that can be derived from BCJ identities between five-point amplitudes.\\
Both algebraic manipulations and numerical evaluations have been carried out by using the {\mathematica} package \sam~\cite{Maitre:2007jq}.


\section{Color-kinematics duality in $d$-dimensions}
In this Section we briefly review the study the C/K-duality for dimensionally regulated amplitudes presented, in the framework of FDF, in \cite{Mastrolia:2015maa}.\\
FDF is a dimensional regularization scheme, first introduced in~\cite{Fazio:2014xea}, which allows a purely four-dimensional representation of the additional degrees of freedom associated to the analytic continuation of the space-time dimension. FDF has been recently applied to the computation of one-loop QCD corrections in \cite{Bobadilla:2015wma,Bobadilla:2016scr}, where the processes $gg \to gg$, $q {\bar q} \to gg$, $gg \to Hg$, $gg \to Hgg$ (in the heavy top
limit) and $gg \to ggg(g)$ were studied.\\
In this formulation, virtual states are associated to massive four-dimensional particles, whose mass acts as regulating parameter.
The four-dimensional degrees of freedom of the gauge bosons are carried by  {\it massive vector bosons} (denoted by $g^{\bullet}$)
of mass $\mu$ and their $(d-4)$-dimensional ones by {\it real scalar particles} ($s^{\bullet}$) of mass
$\mu$.  At the same time, $d$-dimensional fermions of mass $m$ are traded as a {\it tardyonic  Dirac fields} ($q^{\bullet}$) with
mass $m +i \mu \gamma^5$. \\
\label{sec:FDFduality}
 \\
\begin{figure}[h]
	\centering
	\vspace*{0.5cm}
	\includegraphics[scale=0.9]{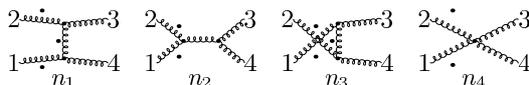}
	\caption{Feynman diagrams for $g^{\bullet}g^{\bullet} \to gg$.}
	\label{fig1}
\end{figure}
In order to show how BCJ identities can be derived taking into account the effects of dimensional regularization, we consider the process $g^{\bullet}(p_1)g^{\bullet}(p_2)\to g(p_3)g(p_4)$, where two generalized gluons, \textit{i.e.} with on-shell momentum $p^2=\mu^2$, produce a final state with two massless ones. The four Feynman diagrams contributing to the amplitude are shown in Fig. \ref{fig1}, where massive particles are indicated with a dot. We anticipate that the following discussion holds for the process $s^{\bullet}s^{\bullet}\to gg$ as well.\\ 
The color factors of the first three diagrams, which involve the exchange of a virtual particle, are, respectively,
\begin{align}
&c_1=\f^{a_2a_3b}\f^{ba_4a_1},& &c_2=\f^{a_1a_2b}\f^{ba_3a_4},& &c_3=\f^{a_1a_3b}\f^{ba_4a_2}.
\end{align}
The four-gluon interaction gives contribution to all of these color structures so that, labelling with $n_i$ the kinematic parts of Feynman graph numerators, it can be decomposed as 
\begin{align}
c_{4}n_{4}=c_{1}n_{1;4}+c_{2}n_{2;4}+c_{3}n_{3;4}.
\end{align}
Therefore, each $n_{i;4}$, conveniently multiplied and divided by the corresponding kinematic pole, can be absorbed into the definition of the numerators of the cubic graphs. As a result, the amplitude is expressed in terms of diagrams involving three-gluon vertices only,
 \begin{align}
 \mathcal{A}_{4}(p_1,p_2,p_3,p_4)=&c_1\frac{n_1}{P_{23}^2-\mu^2}+c_2\frac{n_2}{P_{12}^2}+c_3\frac{n_3}{P_{24}^2-\mu^2},
 \label{amp1}
 \end{align}
 being $P^2_{ij}=(p_i+p_j)^2$. We observe that, in FDF, amplitudes receive contributions from both massless and massive virtual states, as it is evident from the pole structure of the r.h.s. of \eqref{amp1}. \\
 
 The three color factors $c_i$ are related by the Jacobi identity,
\begin{align}
-c_1 +c_2 +c_3=0,
\label{jacobicol}
\end{align}
which allows us, for example by eliminating $c_2$, to rewrite \eqref{amp1} in terms of two color-stripped terms only,
\begin{align}
\mathcal{A}_{4}(p_1,p_2,p_3,p_4)=c_1K_1+c_3K_3,
\label{amp2}
\end{align}
with
\begin{align}
&K_1=\frac{n_1}{P_{23}^2-\mu^2}+\frac{n_2}{P_{12}^2},  &K_3=\frac{n_3}{P_{24}^2-\mu^2}-\frac{n_2}{P_{12}^2}.
\label{Ks}
\end{align}
From the explicit Feynman rules-expression of the numerators $n_i$'s, it can be proven that, when on-shell and transversality conditions ($\epsilon(p_i)\cdot p_i=0$) are imposed, the amplitude satisfies the C/K-duality, \textit{i.e.} the kinematic numerators obey the same Jacobi identity as the color factors,
\begin{align}
-n_1 +n_2 +n_3=0. 
\label{jacobinum}
\end{align}
The set of equations \eqref{Ks} and \eqref{jacobinum} can be conveniently organized into a linear system $\mathbb{A}\,\mathbf{n}=\mathbf{K}$,
\begin{align}
\left( \begin{array}{ccc}
\frac{1}{P_{23}^2-\mu^2} & \;\frac{1}{P_{12}^2} & 0 \\
0 &- \frac{1}{P_{12}^2} &\frac{1}{P_{24}^2-\mu^2} \\
-1 & 1 &1 \end{array} \right)
\left( \begin{array}{c}
n_1  \\
n_2  \\
n_3
\end{array} \right)
=
\left( \begin{array}{c}
K_1  \\
K_3  \\
0
 \end{array} \right).
 \label{sys}
\end{align}
Due to momentum conservation, $P_{12}^2+P_{23}^2+P_{24}^2=2\mu^2$, one can verify that the matrix $\mathbb{A}$ has 
\begin{align}
 \text{rank}(\mathbb{A})=2,
\end{align}
or, equivalently, that a linear relation can be established between its rows,
\begin{align}
(P_{23}^2-\mu^2)\mathbb{A}_1-(P_{24}^2-\mu^2)\mathbb{A}_2+\mathbb{A}_3=0.
\label{rows}
\end{align}
Therefore, because of the consistency condition of the inhomogeneous system \eqref{sys},
\begin{align}
\text{rank}(\mathbb{A})=\text{rank}(\mathbb{A}|\mathbf{K})=2,
\end{align}
a constraint analogous to \eqref{rows} must hold between the elements of the vector $\mathbf{K}$,
\begin{align}
K_3=\frac{P_{23}^2-\mu^2}{P_{24}^2-\mu^2}K_1.
\label{Kid}
\end{align}
Starting from the Feynman diagram expansions \eqref{Ks}, it can be checked that the kinematic factors $K_i$ exactly correspond to two different color-orderings of the amplitude,
\begin{align}
&K_1=A(1,2,3,4), &K_3=A(2,1,3,4),
\end{align}
so that \eqref{Kid} can be rewritten as
\begin{align}
A(2,1,3,4)=\frac{P_{23}^2-\mu^2}{P_{24}^2-\mu^2}A(1,2,3,4).
\label{BCJ4pt1}
\end{align}
With similar considerations, one can verify that
\begin{align}
&A(2,4,1,3)=\frac{P_{12}^2}{P_{24}^2-\mu^2}A(1,2,3,4), & A(2,4,1,3)=\frac{P_{12}^2}{P_{23}^2-\mu^2}A(2,1,3,4).
\label{BCJ4pt2}
\end{align}
Eqs.~\eqref{BCJ4pt1} and \eqref{BCJ4pt2} show how the well known BCJ identities for four-point amplitudes~\cite{Bern:2008qj}, which are formally recovered in the $\mu^2=0$ limit, are extended to FDF tree-level amplitudes, whose massive degrees of freedom keep trace of the effects of dimensional regularization.\\
Analogous relations were proven to hold, at tree-level, for all $2\to 2$ processes involving FDF particles.\\

In general, when moving to higher-point amplitudes,
\begin{align}
\mathcal{A}_{m}(p_1,p_2,\,...\,, p_m)=\sum_{i=1}^{N}\frac{c_in_i}{D_i},
\label{hpamp}
\end{align}
the kinematic numerators obtained in the standard Feynman rules-approach do not satisfy C/K-duality, because of the rising of anomalous terms, which have been shown to  originate from contact interactions.\\
Nevertheless, starting from the set of Feynman rules numerators $n_i$, one can build a dual representation of the amplitude by means of a \textit{generalized gauge transformation}, ~\cite{Bern:2010yg,Monteiro:2011pc,BjerrumBohr:2012mg,Fu:2012uy,Du:2013sha}, \textit{i.e.} a shift of the numerators,
\begin{align}
n_i\to n_i'+\Delta_i,
\end{align}
which leaves the amplitude unchanged,
\begin{align}
\delta\mathcal{A}^{\text{tree}}_{m}(p_1,p_2, ..., p_m)\equiv\sum_{i=1}^{N}\frac{c_i\Delta_i}{D_i}=0,
\label{gengauge}
\end{align}
and re-shuffles the contact terms among numerators, in such a way to restore the C/K-duality.\\
In \cite{Mastrolia:2015maa} a diagrammatic approach was proposed to determine the explicit expressions of the shifts to be performed on the numerators, purely based on the algebraic properties of the higher-point generalization of the linear system
\eqref{sys} and on a systematic way to generate the anomalous terms through the introduction of off-shell currents.\\
In particular, the computation of the rank of the kinematic matrix $\mathbb{A}$ can be used as constructive criterion in order to detect the existing relations between color-ordered amplitudes.\\

As an example, we consider the scattering of two generalized gluons producing three massless ones in the final state, $g^{\bullet}(p_1)g^{\bullet}(p_2)\to g(p_3)g(p_4)g(p_5)$.\\
After absorbing the contributions from four-gluon vertices into the redefinition of the numerators of cubic graphs, the amplitude can be expressed in terms of 15 diagrams, each of them identified by its pole structure, \textit{i.e.} its two internal propagators. 
\\
The color factors associated to these diagrams satisfy a set of 9 independent Jacobi identities of the type
\begin{align}
-c_i+c_j+c_k=0,
\end{align} 
which allow us to express the amplitude in terms of 6 individually gauge invariant terms only,
\begin{align}
\mathcal{A}_5(p_1,p_2,p_3,p_4,p_5)=\sum_{i=1}^6c_iK_i,
\label{amp5g}
\end{align}
with
\begin{align}
c_{1} & =\f^{a_{1}a_{2}b}\f^{ba_{3}c}\f^{ca_{4}a_{5}}, &c_{4}  =\f^{a_{2}a_{3}b}\f^{bca_{1}}\f^{ca_{4}a_{5}},\nonumber \\
c_{2} & =\f^{a_{2}a_{3}b}\f^{ba_{4}c}\f^{ca_{5}a_{1}}, &c_{5}  =\f^{a_{2}bc}\f^{ba_{3}a_{4}}\f^{ca_{5}a_{1}},\nonumber \\
c_{3} & =\f^{a_{1}a_{2}b}\f^{bca_{5}}\f^{ca_{3}a_{4}}, &c_{6}  =\f^{a_{2}a_{5}b}\f^{ba_{3}c}\f^{ca_{4}a_{1}},
\end{align}
and
\begin{align}
&K_1=\frac{n_{1}}{P_{12}^2P_{45}^2}+\frac{n_{12}}{P_{12}^2P_{35}^2}+\frac{n_{13}}{(P_{24}^2-\mu^2)P_{35}^2}-\frac{n_{10}}{(P_{13}^2-\mu^2)(P_{24}^2-\mu^2)}+\frac{n_{15}}{(P_{13}^2-\mu^2)P_{45}^2},\nn
&K_{2}=\frac{n_{2}}{(P_{23}^2-\mu^2)(P_{15}^2-\mu^2)}+\frac{n_{7}}{(P_{14}^2-\mu^2)(P_{23}^2-\mu^2)}-\frac{n_{14}}{(P_{14}^2-\mu^2)P_{35}^2}+\frac{n_{13}}{(P_{24}^2-\mu^2)P_{35}^2}+\frac{n_{11}}{(P_{24}^2-\mu^2)(P_{15}^2-\mu^2)},\nn
&K_3=\frac{n_{3}}{P_{12}^2P_{34}^2}+\frac{n_{9}}{(P_{13}^2-\mu^2)(P_{25}^2-\mu^2)}-\frac{n_{12}}{P_{12}^2P_{35}^2}-\frac{n_{13}}{(P_{24}^2-\mu^2)P_{35}^2}+\frac{n_{10}}{(P_{13}^2-\mu^2)(P_{24}^2-\mu^2)}-\frac{n_{8}}{(P_{25}^2-\mu^2)P_{34}^2},\nn
&K_{4}=\frac{n_{4}}{(P_{23}^2-\mu^2)P_{45}^2}-\frac{n_{7}}{(P_{14}^2-\mu^2)(P_{23}^2-\mu^2)}+\frac{n_{14}}{(P_{14}^2-\mu^2)P_{35}^2}-\frac{n_{13}}{(P_{24}^2-\mu^2)P_{35}^2}+\frac{n_{10}}{(P_{13}^2-\mu^2)(P_{24}^2-\mu^2)}\nn
&\qquad-\frac{n_{15}}{(P_{13}^2-\mu^2)P_{45}^2},\nn
&K_{5}=\frac{n_{5}}{P_{34}^2(P_{15}^2-\mu^2)}-\frac{n_{9}}{(P_{13}^2-\mu^2)(P_{25}^2-\mu^2)}-\frac{n_{10}}{(P_{13}^2-\mu^2)(P_{24}^2-\mu^2)}+\frac{n_{8}}{(P_{25}^2-\mu^2)s_{43}}-\frac{n_{11}}{(P_{24}^2-\mu^2)(P_{15}^2-\mu^2)},\nn
&K_{6}=\frac{n_{6}}{(P_{14}^2-\mu^2)(P_{25}^2-\mu^2)}+\frac{n_{9}}{(P_{13}^2-\mu^2)(P_{25}^2-\mu^2)}+\frac{n_{14}}{(P_{14}^2-\mu^2)P_{35}^2}-\frac{n_{13}}{(P_{24}^2-\mu^2)P_{35}^2}+\frac{n_{10}}{(P_{13}^2-\mu^2)(P_{24}^2-\mu^2)}.
\label{Kfactor5}
\end{align}
The number of distinct gauge invariant contributions to the amplitude, obtained after Jacobi identities are taken into account, corresponds to the number of independent color-ordered amplitudes one gets after imposing Kleiss-Kuijf identity, \cite{Kleiss:1988ne,DelDuca:1999rs}.\\
Since, conversely to the four-point case, the numerators $n_i$'s do not satisfy the same Jacobi identity as the color factors, \eqref{sys} is generalized to a system of 15 equations,
\begin{align}
\mathbb{A}\mathbf{n}=\mathbf{K}+\boldsymbol{\phi},
\end{align}
where
\begin{align}
&\mathbf{n}=(n_1,n_2,\,...\,, n_{15})^{T},\nn
&\mathbf{K}=(\{K_{1},K_2,\,...\,,K_6\},0,0,...\,,0)^{T},\nn
&\boldsymbol\phi=(0,0,\,...\,,0,\{\phi_{[i,j,k]}\})^{T}
\label{vector5pt}
\end{align}
and the elements of the matrix $\mathbb{A}$ take values in
\begin{align}(\mathbb{A})_{ij}\in\{0,\pm 1,\pm (P_{ij}^2)^{-1},\pm(P_{ij}^2-\mu^2)^{-1}\}.
\end{align} 
The anomalous terms $\phi_{[i,j,k]}=-n_i+n_j+n_k$ can be determined recursively starting from four-point off-shell currents.\\
By performing the set of shifts \eqref{gengauge}, one can build an alternative representation of the amplitude, where the $n_i$'s are substituted by a new set of numerators $n_i'$'s satisfying the C/K dual system
\begin{align}
\mathbb{A}\mathbf{n'}=\mathbf{K}.
\end{align}
Because of momentum conservation, the rank of the matrix $\mathbb{A}$ turns out to be non-maximum, $\text{rank}(\mathbb{A})=11$, and the consistency condition
\begin{align}
\text{rank}(\mathbb{A}|\mathbf{K})=11
\end{align}
implies the existence of four linear relations between the kinematic factors $K_i$'s, which can be simply found by determining a complete set of vanishing linear combinations of the rows of $\mathbb{A}$.\\
In this way, we obtain the set of identities
\begin{align}
&P_{45}^2K_1-P_{34}^2K_3-(P_{14}^2-\mu^2)K_6=0,\nn
&P_{12}^2K_1-(P_{23}^2-\mu^2)K_4-(P_{25}^2-\mu^2)K_6=0,\nn
&(P_{15}^2-\mu^2)K_2-P_{45}^2K_4-(P_{25}^2-\mu^2)K_6=0,\nn
&(P_{23}^2-\mu^2)K_2-P_{34}^2K_5+(P_{23}^2+P_{35}^2-\mu^2)K_6=0,
\label{relK5}
\end{align}
which reduce to two the numbers of independent $K_i$'s.\\
 Rather than corresponding to a single partial amplitude, as it was the case at four-point, for higher-multiplicity amplitudes each kinematic factor can be expressed as linear combinations of color-ordered amplitudes. The relations between the $K_i$'s and color-ordered amplitudes can be found either by comparing their expansions in terms of Feynman diagrams or, more conveniently, by first performing the usual color algebra on \eqref{amp5g}, in order to express all $c_i's$ in terms of traces of generators $T^{a_i}$ and then by identifying the combinations of $K_i$'s that multiply each single trace with the corresponding color-ordered amplitude.\\
In this case, it can be shown that
\begin{align}
 &K_1=A_5(1,2,3,4,5)+A_5(1,2,4,3,5)+A_5(1,3,2,4,5),\nn
 &K_2=-A_5(1,4,2,3,5),\nn
 &K_3= A_5(1,3,4,2,5)-A_5(1,2,4,3,5),\nn
 &K_4=A_5(1,4,2,3,5)-A_5(1,3,2,4,5),\nn
 &K_5=-A_5(1,3,4,2,5),\nn
 &K_6= A_5(1,3,4,2,5)+A_5(1,4,2,3,5)+A_5(1,4,3,2,5).
 \label{AtoK}
\end{align}
Therefore, by substituting \eqref{AtoK} in \eqref{relK5}, one can reduce from six to two the number of color-ordered amplitudes, and express all others through the set of relations
\begin{align}
&A_5(1,3,4,2,5)=\frac{-P_{12}^2P_{45}^2A_5(1,2,3,4,5)+(P_{14}^2-\mu^2)(P_{24}^2+P_{25}^2-2\mu^2)A_5(1,4,3,2,5)}{(P_{13}^2-\mu^2)(P_{24}^2-\mu^2)},\nn
&A_5(1,2,4,3,5)=\frac{-(P_{14}^2-\mu^2)(P_{25}^2-\mu^2)A_5(1,4,3,2,5)+P_{45}^2(P_{12}^2+P_{24}^2-\mu^2)A_5(1,2,3,4,5)}{P_{35}^2(P_{24}^2-\mu^2)},\nn
&A_5(1,4,2,3,5)=\frac{-P_{12}^2P_{45}^2A_5(1,2,3,4,5)+(P_{25}^2-\mu^2)(P_{14}^2+P_{25}^2-2\mu^2)A_5(1,4,3,2,5)}{P_{35}^2(P_{24}^2-\mu^2)},\nn
&A_5(1,3,2,4,5)=\frac{-(P_{14}^2-\mu^2)(P_{25}^2-\mu^2)A_5(1,4,3,2,5)+P_{12}^2(P_{24}^2+P_{45}^2-\mu^2)A_5(1,2,3,4,5)}{(P_{13}^2-\mu^2)(P_{24}^2-\mu^2)}.
\label{BCJ5pt}
\end{align}
Identities involving other color-ordered amplitudes can be obtained by making use of Kleiss-Kuijf identities, such as,
\begin{align}
&A_5(1,2,3,4,5)+A_5(1,2,3,5,4)+A_5(1,2,4,3,5)+A_5(1,4,2,3,5)=0,\nn
\end{align}
which, substituted in \eqref{BCJ5pt}, gives,
\begin{align}
&A_5(1,2,4,3,5)=\frac{(P_{14}^2+P_{45}^2-\mu^2)A_5(1,2,3,4,5)+(P_{14}^2-\mu^2)A_5(1,2,3,5,4)}{(P_{24}^2-\mu^2)}.
\label{BCJ5pt1}
\end{align}
Again, this set of identities corresponds to the FDF extension of the BCJ relations between four-dimensional color-ordered amplitudes~\cite{Bern:2008qj}, which are recovered by setting $\mu^2=0$. The very same identities are satisfied by the color-ordered amplitudes where the generalized gluons in the initial state are replaced by massive scalars, $s^{\bullet}s^{\bullet}\to ggg$, and similar relations have been proven in \cite{Mastrolia:2015maa} for five-point amplitudes involving quarks, namely $g^{\bullet}g^{\bullet}(s^{\bullet}s^{\bullet})\to q\bar{q}g$. Moreover, the method we have summarized can find a straightforward generalization to higher-multiplicities.\\

In the following Section, we will show how FDF formulation of BCJ identities for $d$-dimensional tree-levels, such as \eqref{BCJ4pt1} and \eqref{BCJ5pt1}, can be used in order to determine coefficient relations for full $d$-dimensional one-loop amplitudes, including both cut constructible part and rational terms.

\section{Coefficient relations for one-loop amplitudes in $d$-dimensions}
\label{sec:coef-relations}
Since the introduction of generalized unitarity~\cite{Bern:1994zx, Britto:2004nc} and complex kinematics for on-shell particles~\cite{Cachazo:2004kj,Britto:2004ap}, the study of analyticity and factorization properties of scattering amplitudes has turned into an extremely powerful tool for their computation.\\ Relying on the decomposition of any amplitude as a linear combination of master integrals (MI's)~\cite{Passarino:1978jh,vanOldenborgh:1989wn}, the basic idea of unitarity based methods consists in extracting the coefficients of the MI's by matching multiple cuts of the amplitude with the cuts of the MI's themselves.\\
In this framework, the \textit{integrand reduction method}, first introduced for one-loop amplitudes in \cite{Ossola:2006us} and \cite{Ellis:2007br}, in four- and $d$-dimensions respectively, and more recently extended to multi-loop case, \cite{Mastrolia:2011pr,Badger:2012dp,Zhang:2012ce,Mastrolia:2012an,Mastrolia:2013kca}, exploits the knowledge of the algebraic structure of Feynman integrands, which allows to decompose each numerators as a combination of products of denominators with polynomial coefficients, in order to reach the decomposition of scattering amplitudes in terms of MI's.\\
At one-loop, if we split the $d=4-2\epsilon$ dimensional loop momentum $\bar{q}^{\alpha}$ into its four-dimensional part $q^{\alpha}$ and a vector $\mu^{\alpha}$ belonging to the $-2\epsilon$-subspace,
\begin{align}
\slashed{\bar{q}}=\slashed{q}+\slashed{\mu},\qquad \bar{q}^2=q^2-\mu^2,
\end{align}
we can write an arbitrary one-loop $n$-point color-ordered amplitude as
\begin{align}
A_n^{1\text{-loop}}& =\int d^d\bar q \frac{\mathcal{N}(q,\mu^2)}{D_0D_1\,...\,D_{n-1}},
\label{amp}
\end{align}
with
\begin{align}
D_i & =(\bar{q}+p_i)^2-m_i^2=(q+p_i)^2-m_i^2-\mu^2.
\end{align}
The integrand reduction algorithm allows us to write the numerators $\mathcal{N}(q,\mu^2)$ in terms of denominators and, consequently, to obtain a decomposition of the integrand of the type
\begin{align}
\frac{\mathcal\mathcal{N}(q,\mu^2)}{D_0D_1\,...\,D_{n-1}}=&\sum_{i\ll m}^{n-1}\frac{\Delta_{ijklm}(q,\mu^2)}{D_iD_jD_kD_lD_m}+\sum_{i\ll l}^{n-1}\frac{\Delta_{ijkl}(q,\mu^2)}{D_iD_jD_kD_l}+\sum_{i\ll k}^{n-1}\frac{\Delta_{ijk}(q,\mu^2)}{D_iD_jD_k}\nn
&+\sum_{i\l j }^{n-1}\frac{\Delta_{ij}(q,\mu^2)}{D_iD_j}+\sum_{i}^{n-1}\frac{\Delta_{i}(q,\mu^2)}{D_i},
\label{integranddeco}
\end{align}
where $i\ll m$ indicates lexicographic ordering. The functions $\Delta_{i\cdots k}(q,\mu^2)$, called residues, are polynomials in $\mu^2$ and in the components $\{x_i\}$ of $q$, which, according to the cut $D_i=D_j=D_k=\cdots=0$ under consideration, is decomposed with respect to a suitable basis of four-dimensional vectors $\mathcal{E}^{(i\cdots k)}$ (see, for instance, \cite{Mastrolia:2012bu}).\\
The parametric expression $\Delta_{i\cdots k}(q,\mu^2)$ is process-independent and, for renormalizable theories~\cite{Ossola:2006us,Ellis:2007br,Zhang:2012ce}, is given by
\begin{align}
\Delta_{ijklm}&=c\mu^2,\nn
\Delta_{ijkl}&=c_0+c_1x_{4,v}+c_2\mu^2+c_3x_{4,v}\mu^2+c_4\mu^4,\nn
\Delta_{ijk}&=c_{0,0}+c_{1,0}^{+}x_4+c_{2,0}^{+}x_4^2+c_{3,0}^{+}x_4^3+c_{1,0}^{-}x_3+c_{2,0}^{-}x_3^2+c_{3,0}^{-}x_3^3+c_{0,2}\mu^2+c_{1,2}^{+}x_4\mu^2+c_{1,2}^{-}x_3\mu^2,\nn
\Delta_{ij}&=c_{0,0,0}+c_{0,1,0}x_1+c_{0,2,0}x_1^2+c_{1,0,0}^{+}x_4+c_{2,0,0}^{+}x_4^2+c_{1,0,0}^{-}x_3+c_{2,0,0}^{-}x_3^2+c_{1,1,0}^{+}x_1x_4+c_{1,1,0}^{-}x_1x_3+c_{0,0,2}\mu^2,\nn
\Delta{i}&=c_{0,0,0,0}+c_{0,1,0,0}x_1+c_{0,0,1,0}x_2+c^{-}_{1,0,0,0}x_3+c^{+}_{1,0,0,0}x_4,
\end{align}
\label{res}
where, for each coefficient, a superscript labelling the specific cut is understood, $c_l=c_l^{(i\cdots k)}$.\\
As a consequence of \eqref{res}, by neglecting all \textit{spurious} terms, which vanish upon integration, the amplitude \eqref{amp} can be written in terms of MI's
\begin{align}
I_{i\cdots k}[\alpha]=\int d^d\bar{q}\frac{\alpha}{D_i\cdots D_k}
\end{align}
and of the coefficients of the residues as
\begin{align}
A_n^{1\text{-loop}}=&\sum_{i\ll l}^{n-1}\left[c^{(ijkl)}_0I_{ijkl}[1]+c^{(ijkl)}_4I_{ijkl}[\mu^4]\right]+\sum_{i\ll k}^{n-1}\left[c^{(ijk)}_{0,0}I_{ijk}[1]+c^{(ijk)}_{0,2}I_{ijk}[\mu^2]\right]+\nn&\sum_{i\ll j}^{n-1}\left[c^{(ij)}_{0,0,0}I_{ij}[1]+c^{(ij)}_{0,1,0}I_{ij}[(q+p_i)\cdot e_2]+c^{(ij)}_{0,2,0}I_{ij}[((q+p_i)\cdot e_2)^2]+c^{(ij)}_{0,0,2}I_{ij}[\mu^2]\right]
+\sum_{i}^{n-1}c^{(i)}_{0,0,0,0}I_{i}[1].
\label{integraldeco}
\end{align}
In the original top-down formulation of the algorithm \cite{Ossola:2006us,Mastrolia:2010nb,Mastrolia:2008jb}, all coefficients of the integrand decomposition \eqref{integranddeco} are computed by sampling
the numerator of the amplitude, after all non-vanishing contribution to higher-point residues have been subtracted, on a finite set of on-shell solutions of the multiple cuts.\\
Alternatively, starting from the techniques presented in \cite{Forde:2007mi,Badger:2008cm}, it has been shown in \cite{Mastrolia:2012bu} that, by performing a suitable Laurent expansion of the integrand, evaluated on the cut, with respect to one of the components of the loop momenta which are left unconstrained by the on-shell conditions, one can determine the unknown coefficients of the integrand reduction by comparison with the ones of the Laurent expansion itself.\\

A full color-dressed amplitude is obtained as a combination color-ordered amplitudes, multiplied for the corresponding color structure. For instance, in the pure-gluon case, we have~\cite{Bern:1990ux,Bern:1994zx}
\begin{align}
\mathcal{A}_n^{1\text{-loop}}=&g^n\sum_{c=1}^{[n/2]+1}\sum_{\sigma\in S_n/S_{n;c}}\text{Gr}_{n;c}(\sigma)A_{n;c}^{1\text{-loop}}(\sigma),\nn
&\text{Gr}_{n;1}(\sigma)=N_c\text{Tr}(T^{a_{\sigma(1)}}\cdots T^{a_{\sigma(n)}}),\nn
&\text{Gr}_{n;c}(\sigma)=N_c\text{Tr}(T^{a_{\sigma(1)}}\cdots T^{a_{\sigma(c-1)}})\text{Tr}(T^{a_{\sigma(c)}}\cdots T^{a_{\sigma(c-1)}}),\quad c>1.
\end{align}
Although it is sufficient to consider leading color contributions $A_{n;1}(\sigma)\equiv A_{n}(\sigma)$, since amplitudes associated to subleading colors can be obtained as a sum over permutations of $A_{n}(\sigma)$'s, one should, in principle, fit the coefficients of the residues \eqref{res} for each color-ordering.\\ However, C/K-duality satisfied by tree-level amplitudes, in which the integrand factorizes when evaluated on unitarity cuts, can be used to determine relations between coefficients of residues which differ from the ordering of external particles, and thus to reduce the total number of coefficients to be individually computed.\\

In the following, we recall the extraction of coefficients via Laurent expansion, for which we refer to \cite{Mastrolia:2012bu} and \cite{Peraro:2014cba}, and we make use of the $d$-dimensional BCJ identities presented in Section~\ref{sec:FDFduality} in order to determine the full set of relations between integral coefficients. As we will explicitly show, these identities holds separately for both independent cut solutions that must be averaged in the extraction of the integral coefficients.\\ For sake of simplicity, we derive relations between integral coefficient that can be obtained starting from  BCJ identities at four-point only and we collect in \ref{apd} the set of relations that follow from C/K-duality for five-point amplitudes.\\
We expect similar results to hold even when BCJ identities for higher-multiplicity amplitudes are taken into account but we leave this generalization to future studies.
For this reason, we will not discuss relations between tadpoles coefficients, which would at least require BCJ identities for six-point tree-levels.

\subsection{Relations for pentagon coefficients}
\label{subpenta}
\begin{figure}[h]
	\centering
	\vspace*{0.5cm}
	\includegraphics[scale=1.1]{fig2.epsi}
	\caption{Pentagon topologies for the cuts $	C_{12|3\ldots k|\left(k+1\right)\ldots l|\left(l+1\right)\ldots m|\left(m+1\right)\ldots n}$ and $	C_{21|3\ldots k|\left(k+1\right)\ldots l|\left(l+1\right)\ldots m|\left(m+1\right)\ldots n}$.}
	\label{figs/5sP.21}
\end{figure}
The solutions of the quintuple cut $D_{i}=D_{j}=D_{k}=D_{l}=D_{m}=0$ can be parametrized as
\begin{align}
l_{+}^{\left(ijklm\right)\nu}= & p_{i}+x_{1}e_{1}^{\left(ijklm\right)\nu}+x_{2}e_{2}^{\left(ijklm\right)\nu}+x_{3}e_{3}^{\left(ijklm\right)\nu}+\frac{x_{4}+\mu^{2}}{x_{3}}e_{4}^{\left(ijklm\right)\nu},\\
l_{-}^{\left(ijklm\right)\nu}= & p_{i}+x_{1}e_{1}^{\left(ijklm\right)\nu}+x_{2}e_{2}^{\left(ijklm\right)\nu}+x_{3}e_{4}^{\left(ijklm\right)\nu}+\frac{x_{4}+\mu^{2}}{x_{3}}e_{3}^{\left(ijklm\right)\nu},
\end{align}
where the full set of parameters $x_{1},x_{2},x_{3},x_{4}$ and $\mu^{2}$ is fixed by the cut conditions. 
The single pentagon coefficient appearing in \eqref{integranddeco} can be computed evaluating the integrand on the two on-shell solutions,
\begin{align}
C_{i|j|k|l|m}^{\pm}= & \frac{N_{\pm}}{\prod_{h\ne i,j,k,l,m}D_{h,\pm}}=c^{(ijklm)\pm}\mu^2.
\label{pentacoef}
\end{align}
In order to see how BCJ identities for tree-level amplitudes can be used to relate different pentagon coefficients, let us consider the pentagon contributions shown in Fig.~\ref{figs/5sP.21}, which share the same cut solutions. In addition, since these two pentagons differ in the ordering of the external particles $p_1$ and $p_2$ only, they can be obtained as the product of the same tree-level amplitudes, with the only exception of the color-ordering of the four-point amplitude involving $p_1$ and $p_2$.\\
More precisely, for the ordering $\{1,2,\, ...\, ,n\}$ we have
	\begin{align}
	C^{\pm}_{12|3\ldots k|\left(k+1\right)\ldots l|\left(l+1\right)\ldots m|\left(m+1\right)\ldots n}= & A_{4}^{\text{tree}}\left(-l^{\pm}_{1},1,2,l^{\pm}_{3}\right)A_{k}^{\text{tree}}\left(-l^{\pm}_{3},P_{3\cdots k},l^{\pm}_{k+1}\right)A_{l-k+2}^{\text{tree}}\left(-l^{\pm}_{k+1},P_{k+1\ldots,l^{\pm}},l^{\pm}_{l}\right)\nonumber \\
	& \qquad\times A_{m-l+2}^{\text{tree}}\left(-l_{l+1}^{\pm},P_{l^{\pm}+1\ldots,m},l^{\pm}_{m}\right)A_{n-m+2}^{\text{tree}}\left(-l^{\pm}_{m+1},P_{m+1\ldots,n},l^{\pm}_{1}\right)
	\end{align}
and $C^{\pm}_{21|3\ldots k|\left(k+1\right)\ldots l|\left(l+1\right)\ldots m|\left(m+1\right)\ldots n}$ is obtained just by changing $1\leftrightarrow 2$.\\

The tree-level amplitudes $A_{4}^{\text{tree}}\left(-l^{\pm}_{1},1,2,l^{\pm}_{3}\right)$ and  $A_{4}^{\text{tree}}\left(-l^{\pm}_{1},2,1,l^{\pm}_{3}\right)$ are related by the $d$-dimensional BCJ identity \eqref{BCJ4pt2}, 
\begin{align}
A_{4}^{\text{tree}}(-l^{\pm}_{1},2,1,l^{\pm}_{3})=\frac{P_{l^{\pm}_{3}2}^{2}-\mu^{2}}{P_{-l^{\pm}_{1}2}^{2}-\mu^{2}}A_{4}^{\text{tree}}(-l^{\pm}_{1},1,2,l^{\pm}_{3}),
\label{eq:bcjt4}
\end{align}
which, substituted into the expression of $C^{\pm}_{21|3\ldots k|\left(k+1\right)\ldots l|\left(l+1\right)\ldots m|\left(m+1\right)\ldots n}$, allow us to identify
	\begin{align}
	C^{\pm}_{21|3\ldots k|\left(k+1\right)\ldots l|\left(l+1\right)\ldots m|\left(m+1\right)\ldots n}=\frac{P_{l^{\pm}_{3}2}^{2}-\mu^{2}}{P_{-l^{\pm}_{1}2}^{2}-\mu^{2}}C^{\pm}_{12|3\ldots k|\left(k+1\right)\ldots l|\left(l+1\right)\ldots m|\left(m+1\right)\ldots n}.
	\label{eq:pentaC21}
	\end{align}
The ratio of the two propagators appearing in \eqref{eq:pentaC21} produces the same constant for both cut solutions,
	\begin{align}
	\frac{P_{l^{\pm}_{3}2}^{2}-\mu^{2}}{P_{-l^{\pm}_{1}2}^{2}-\mu^{2}} & =\alpha,
	\label{ratio5pt}
	\end{align}
	so that, making use of \eqref{pentacoef}, \eqref{eq:pentaC21} becomes 
	\begin{align}
	c^{\left(21|\ldots\right)\pm} & =\alpha c^{\left(12|\ldots\right)\pm}.
	\label{cf5}
	\end{align}
Therefore, as byproduct of BCJ identities at tree-level, the knowledge of a single pentagon coefficient immediately allow us to obtain the other one.
\subsection{Relations for box coefficients}
\label{subbox}

\begin{figure}[h]
	\centering
	\vspace*{0.5cm}
	\includegraphics[scale=1.1]{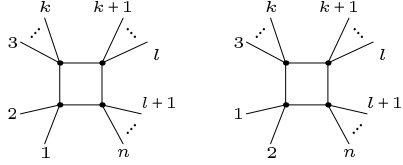}
	\caption{Box topologies for the cuts $C_{12|3\ldots k|k+1\ldots l|l+1\ldots n}$ and $C_{21|3\ldots k|k+1\ldots l|l+1\ldots n}$.}
	\label{figs/5sQ.21}
\end{figure}
Next we consider the quadrupole cut $D_{i}=D_{j}=D_{k}=D_{l}=0$, whose solutions are parametrized as
\begin{align}
& l_{\pm}^{\left(ijkl\right)\nu}=-p_{i}^{\nu}+x_{1}\,e_{1}^{\left(ijkl\right)\nu}+x_{2}\,e_{2}^{\left(ijkl\right)\nu}+x_{v}\,v^{\left(ijkl\right)\nu}\pm u\,v_{\perp}^{\left(ijkl\right)\nu}, &  & u=\sqrt{a_{\perp}+\frac{\mu^{2}}{v_{\perp}^{2}}},
\end{align}
being $x_{1},x_{2},x_{v}$ and $a_{\perp}$ coefficients fixed by the cut-conditions.\\
The two non-spurious coefficients can be extracted in the $\mu^2\to0$ and $\mu^2\to\infty$ limits,
\begin{subequations}
	\begin{align}
	\label{exbox1}
	&C_{i|j|k|l}^{\pm}=\frac{N_{\pm}}{\prod_{h\ne i,j,k,l}D_{h,\pm}}\bigg|_{\mu^{2}\to0}  =c_{0}^{(ijkl)\pm},\\
	\label{exbox2}
	&C_{i|j|k|l}^{\pm}=\frac{N_{\pm}}{\prod_{h\ne i,j,k,l}D_{h,\pm}}\bigg|_{\mu^{2}\to\infty}  =c_{4}^{(ijkl)\pm}\mu^{4}+\mathcal{O}\left(\mu^{3}\right),
	\end{align}
\end{subequations}
and the box contribution to the amplitude \eqref{integraldeco} is obtained by averaging over the two cut solutions,
\begin{align}
A_{n}^{1\text{-loop}}\bigg|_{\text{box}} & =\frac{1}{2}\left(c_{0}^{(ijkl)+}+c_{0}^{(ijkl)-}\right)I_{ijkl}\left[1\right]+c_{4}^{(ijkl)}\,I_{(ijkl)}\left[\mu^{4}\right],
\end{align}
where we used $c_{4}^{(ijkl)}\equiv c_{4}^{(ijkl)+}=c_{4}^{(ijkl)-}$.\\
Analogously to the pentagon case, we consider two box topologies differing just from the ordering of the external particles $p_1$ and $p_2$, as depicted in Fig.~\ref{figs/5sQ.21}.\\
Evaluating the integrand associated to the ordering $\{1,2,\,...\, ,n\}$ on the on-shell solutions, it factorizes into
\begin{align}
	C^{\pm}_{12|3\ldots k|\left(k+1\right)\ldots l|\left(l+1\right)\ldots n} & =A_{4}^{\text{tree}}\left(-l^{\pm}_{1},1,2,l^{\pm}_{3}\right)A_{k}^{\text{tree}}\left(-l^{\pm}_{3},P_{3\cdots k},l^{\pm}_{k+1}\right)A_{l-k+2}^{\text{tree}}\left(-l^{\pm}_{k+1},P_{k+1\ldots,l},l^{\pm}_{l+1}\right)\nn
	&\times A_{n-l+2}^{\text{tree}}\left(-l^{\pm}_{l+1},P_{l+1\ldots,n},l^{\pm}_{1}\right),
\end{align}
and the expression of $C^{\pm}_{21|3\ldots k|\left(k+1\right)\ldots l|\left(l+1\right)\ldots n}$ in terms of tree-amplitudes can be obtained  by changing $1\leftrightarrow 2$.\\
Therefore, thanks to the identity between tree-level amplitudes \eqref{eq:bcjt4}, we can write
	\begin{align}
	& C^{\pm}_{21|3\ldots k|\left(k+1\right)\ldots l|\left(l+1\right)\ldots n}=\frac{P_{l^{\pm}_{3}2}^{2}-\mu^{2}}{P_{-l^{\pm}_{1}2}^{2}-\mu^{2}}C^{\pm}_{12|3\ldots k|\left(k+1\right)\ldots l|\left(l+1\right)\ldots n}.
	\label{C21box}
	\end{align}
It can be verified that the ratio of propagators, when evaluated on the cut solutions, converges to a constant in both $\mu^{2}\to0$ and $\mu^{2}\to\infty$
	limits
	\begin{align}
	&\left.\frac{P_{l^{\pm}_{3}2}^{2}-\mu^{2}}{P_{-l^{\pm}_{1}2}^{2}-\mu^{2}}\right|_{\mu^{2}\to0}  =\alpha_{0}^{\pm},\\
	&\left.\frac{P_{l^{\pm}_{3}2}^{2}-\mu^{2}}{P_{-l^{\pm}_{1}2}^{2}-\mu^{2}}\right|_{\mu^{2}\to\infty} =\alpha_{4}^{\pm}+\mathcal{O}\left(\frac{1}{\mu}\right),
	\end{align}
	so that, by evaluating both sides of \eqref{C21box} in the two limits, we can trivially obtain the contributions from $C_{21|3\ldots k|\left(k+1\right)\ldots l|\left(l+1\right)\ldots n}$, once $C_{12|3\ldots k|\left(k+1\right)\ldots l|\left(l+1\right)\ldots n}$ is known, 
	\begin{align}
	& c_{i}^{\left(21|\ldots\right)\pm}=\alpha_{i}^{\pm}c_{i}^{\left(12|\ldots\right)\pm},\qquad i=0,4.
	\label{cf4}
	\end{align}
\subsection{Relations for triangle coefficients}
\label{subtri}
\begin{figure}[h]
	\centering
	\vspace*{0.5cm}
	\includegraphics[scale=1.1]{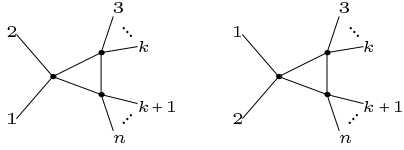}
\caption{Triangle topologies for the cuts $C_{12|3\ldots k|\left(k+1\right)\ldots n}$ and $C_{21|3\ldots k|\left(k+1\right)\ldots n}$.}
\label{figs/5sT.12}
\end{figure}
The solutions of the triple cut $D_{i}=D_{j}=D_{k}=0$ can be parametrized in terms of $\mu^2$ and one free parameter $t$ as
\begin{align}
	l_{+}^{\left(ijk\right)\nu}= & p_{i}+x_{1}e_{1}^{\left(ijk\right)\nu}+x_{2}e_{2}^{\left(ijk\right)\nu}+t\,e_{3}^{\left(ijk\right)\nu}+\frac{x_{3}+\mu^{2}}{t}e_{4}^{\left(ijk\right)\nu},\nn
	l_{-}^{\left(ijk\right)\nu}= & p_{i}+x_{1}e_{1}^{\left(ijk\right)\nu}+x_{2}e_{2}^{\left(ijk\right)\nu}+\frac{x_{3}+\mu^{2}}{t}e_{3}^{\left(ijk\right)\nu}+t\,e_{4}^{\left(ijk\right)\nu},
	\label{3cutsol}
\end{align}
where the coefficients $x_{1}$, $x_{2}$ and $x_{3}$ are fixed by the cut conditions.\\
Starting from the expansion of the integrand in the large-$t$ limit,
\begin{align}
	C_{i|j|k}^{\pm}\left(t,\mu^{2}\right)&=\frac{N_{\pm}}{\prod_{h\ne i,j,k}D_{h,\pm}}\bigg|_{t\to\infty}=\sum_{m=0}^{3}c_{m,0}^{\left(ijk\right)\pm}t^{m}+\mu^{2}\sum_{m=0}^{1}c_{m,2}^{\left(ijk\right)\pm}t^{m},
	\label{eq:3cute}
\end{align}
the triangle contribution to the one-loop amplitude \eqref{integraldeco} is then obtained by averaging on the two solutions \eqref{3cutsol}.
\begin{align}
	 A_{n}^{1\text{-loop}}\big|_{\text{triangle}} & =\frac{1}{2}\left(c_{0,0}^{+}+c_{0,0}^{-}\right)I_{3}\left[1\right]+\frac{1}{2}\left(c_{0,2}^{+}+c_{0,2}^{-}\right)I_{3}\left[\mu^{2}\right].
\end{align}
C/K-duality for tree-level amplitudes can be used to relate all coefficients in the expansions \eqref{eq:3cute} for different triangles. We consider the two triangle contributions depicted in Fig.~\ref{figs/5sT.12}. When evaluated on the on-shell solutions, the triangle with ordering $\{1,2,\,...\,, n\}$ factorizes into
\begin{subequations}
	\begin{align}
	C^{\pm}_{12|3\ldots k|\left(k+1\right)\ldots n} & =A_{4}^{\text{tree}}\left(-l^{\pm}_{1},1,2,l^{\pm}_{3}\right)A_{k}^{\text{tree}}\left(-l^{\pm}_{3},P_{3\cdots k},l^{\pm}_{k+1}\right)A_{n-k+2}^{\text{tree}}\left(-l^{\pm}_{k+1},P_{k+1\ldots,n},l^{\pm}_{1}\right)
	\end{align}
\end{subequations}
and the analogous expression for $C^{\pm}_{21|3\ldots k|\left(k+1\right)\ldots n}$ is obtained by changing $1\leftrightarrow 2$.\\
As for the previous cases, we can make use of the BCJ identity \eqref{eq:bcjt4} in order to establish a relation between $C^{\pm}_{21|3\ldots k|\left(k+1\right)\ldots n}$ and $C^{\pm}_{12|3\ldots k|\left(k+1\right)\ldots n}$,
\begin{align}
C^{\pm}_{21|3\ldots k|\left(k+1\right)\ldots n} =\frac{P_{l^{\pm}_{3}2}^{2}-\mu^{2}}{P_{-l^{\pm}_{1}2}^{2}-\mu^{2}}C_{12|3\ldots k|\left(k+1\right)\ldots n}.
\label{eq:bcj4pt3c}
\end{align}
According to the expansion \eqref{eq:3cute}, both $C^{\pm}_{21|3\ldots k|\left(k+1\right)\ldots n}$ and $C^{\pm}_{12|3\ldots k|\left(k+1\right)\ldots n}$ can be parametrized as
\begin{align}
C_{12|3\ldots k|\left(k+1\right)\ldots n}^{\pm} & =\sum_{m=0}^{3}c_{m,0}^{\left(12|\ldots\right)\pm}t^{m}+\mu^{2}\sum_{m=0}^{1}c_{m,2}^{\left(12|\ldots\right)\pm}t^{m},\nn
C_{21|3\ldots k|\left(k+1\right)\ldots n}^{\pm} & =\sum_{m=0}^{3}c_{m,0}^{\left(21|\ldots\right)\pm}t^{m}+\mu^{2}\sum_{m=0}^{1}c_{m,2}^{\left(21|\ldots\right)\pm}t^{m}.
\label{eq:12,21}
\end{align}
Hence, we can consider the large-$t$ limit of the ratio of the two propagators evaluated on the cut-solution, which is found in the form
	\begin{align}
	\left.\frac{P_{l^{\pm}_{3}2}^{2}-\mu^{2}}{P_{-l^{\pm}_{1}2}^{2}-\mu^{2}}\right|_{t\to\infty} & =\sum_{m=-3}^{0}\alpha_{m,0}^{\pm}t^{m}+\mu^{2}\sum_{m=-3}^{-2}\alpha_{m,2}^{\pm}t^{m}+\mathcal{O}\left(\frac{1}{t^{4}}\right),
	\label{eq:rat3c}
	\end{align}
 and, by plugging the expansions \eqref{eq:12,21} and \eqref{eq:rat3c} into \eqref{eq:bcj4pt3c} we can compare each monomial between the two sides and obtain the set of relations
\begin{align}
& c_{m,0}^{\left(21|\ldots\right)\pm}=\sum_{l=0}^{3-m}\alpha_{-l,0}^{\pm}\,c_{l+m,0}^{\left(12|\ldots\right)\pm}, &  & c_{m,2}^{\left(21|\ldots\right)\pm}=\sum_{l=0}^{1-m}\left(\alpha_{-l-2,2}^{\pm}\,c_{l+m+2,0}^{\left(12|\ldots\right)\pm}+\alpha_{-l0}^{\pm}\,c_{l+m,2}^{\left(12|\ldots\right)\pm}\right).
\label{reltriangle}
\end{align}
Eqs.~\eqref{reltriangle} show that $C^{\pm}_{21|3\ldots k|\left(k+1\right)\ldots n}$ can be fully reconstructed from the knowledge of $C^{\pm}_{12|3\ldots k|\left(k+1\right)\ldots n}$.\\

\subsection{Relations for bubble coefficients}
\label{subbubble}
\begin{figure}[h]
	\centering
	\vspace*{0.5cm}
	\includegraphics[scale=1.1]{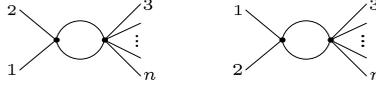}
	\caption{Bubble topologies for the cuts $C_{12|3\ldots n}$ and $C_{21|3\ldots n}$.}
	\label{figs/5sD.21.tex}
\end{figure}
Finally, we consider the double cut $D_{i}=D_{j}=0$, whose solutions are parametrized as 
	\begin{align}
	l_{+}^{\left(ij\right)\nu} & =-p_{i}^{\nu}+y\,e_{1}^{\left(ij\right)\nu}+\left(a_{0}+y\,a_{1}\right)e_{2}^{\left(ij\right)\nu}+t\,e_{3}^{\left(ij\right)\nu}+\frac{\mu^{2}+b_{0}+b_{1}y+b_{2}y^{2}}{t}e_{4}^{\left(ij\right)\nu},\nn
	l_{-}^{\left(ij\right)\nu} & =-p_{i}^{\nu}+y\,e_{1}^{\left(ij\right)\nu}+\left(a_{0}+y\,a_{1}\right)e_{2}^{\left(ij\right)\nu}+\frac{\mu^{2}+b_{0}+b_{1}y+b_{2}y^{2}}{t}e_{3}^{\left(ij\right)\nu}+t\,e_{4}^{\left(ij\right)\nu},
	\end{align}
where $a_{i}$ and $b_{i}$ are kinematic factors fixed by the cut conditions, whereas $t$ and $y$ are free parameters.\\
The bubble coefficients are extracted from the large-$t$ expansion, which returns
\begin{align}
C_{i|j}^{\pm}\left(t,y,\mu^{2}\right)=\left.\frac{N_{\pm}}{\prod_{h\ne i,j}D_{h,\pm}}-\sum_{k\ne i,j}^{n-1}\frac{\Delta_{ijk,\pm}^{R}}{D_{k,+}}\right|_{t\to\infty} & =\sum_{l=0}^{2}\sum_{m=0}^{2-l}c_{l,m,0}^{\left(ij\right)\pm}\,t^{l}\,y^{m}+\mu^{2}c_{0,0,2}^{\left(ij\right)\pm}\label{eq:2cut}
\end{align}
where $\Delta_{ijk,\pm}^{R}$ are defined in~\cite{Mastrolia:2012bu} and are needed in order to subtract \textit{spurious} contributions originating from triangle coefficients.\\
The bubble contribution to the amplitude \eqref{integraldeco} is
\begin{align}
A_{n}^{1\text{-loop}}\bigg|_{\text{bubble}}= & c_{0,0,0}^{\left(ij\right)}I_{ij}\left[1\right]+c_{0,1,0}I_{ij}\left[(q+p_i)\cdot e_2\right]+c_{0,2,0}I_{ij}\left[((q+p_i)\cdot e_2)^{2}\right]+c_{0,0,2}^{\left(ij\right)}I_{ij}\left[\mu^{2}\right],
\end{align}
 where we dropped the \textquotedblleft$\pm$\textquotedblright $\,$  label, since the coefficients appearing in the r.h.s. turn out to be the identical for the two solutions.\\ 

As usual, in order to show the role of C/K-duality in the reduction of the number of coefficients to be actually computed, we consider two bubble contributions
 differing by the ordering of the external particles $p_1$ and $p_2$, as illustrated in Fig.~\ref{figs/5sD.21.tex}.\\
 The two coefficients are given by
	\begin{align}
	C^{\pm}_{12|3\ldots n} & =A_{4}^{\text{tree}}\left(-l^{\pm}_{1},1,2,l^{\pm}_{3}\right)A_{n}^{\text{tree}}\left(-l^{\pm}_{3},P_{3\cdots n},l^{\pm}_{1}\right),\nn
	C^{\pm}_{21|3\ldots n} & =A_{4}^{\text{tree}}\left(-l^{\pm}_{1},2,1,l^{\pm}_{3}\right)A_{n}^{\text{tree}}\left(-l^{\pm}_{3},P_{3\cdots n},l^{\pm}_{1}\right)
	\end{align}
	and, using \eqref{eq:bcjt4} to relate $A_{4}^{\text{tree}}\left(-l^{\pm}_{1},1,2,l^{\pm}_{3}\right)$ and $A_{4}^{\text{tree}}\left(-l^{\pm}_{1},2,1,l^{\pm}_{3}\right)$ we obtain
	\begin{align}
	C^{\pm}_{21|3\ldots n}  =\frac{P_{l^{\pm}_{3}2}^{2}-\mu^{2}}{P_{-l^{\pm}_{1}2}^{2}-\mu^{2}}C^{\pm}_{12|3\ldots n}.
	\label{eq:bcj4pt2c}
	\end{align}
    The ratio of propagators in the large-$t$ limit is parametrized as
	\begin{align}
	\left.\frac{P_{l^{\pm}_{3}2}^{2}-\mu^{2}}{P_{-l^{\pm}_{1}2}^{2}-\mu^{2}}\right|_{t\to\infty} & =\sum_{l=-2}^{0}\sum_{m=0}^{-l}\alpha_{l,m,0}\,t^{l}\,y^{m}+\frac{\mu^{2}}{t^{2}}\alpha_{-2,0,2}+\mathcal{O}\left(\frac{1}{t^3}\right),
	\label{eq:rat2c}
	\end{align}
	so that, by plugging in \eqref{eq:bcj4pt2c} the expansions
	\begin{align}
	C_{12|3\ldots n}^{\pm}= & \sum_{l=0}^{2}\sum_{m=0}^{2-l}c_{l,m,0}^{\left(12|\ldots\right)\pm}\,t^{l}\,y^{m}+\mu^{2}c_{0,0,2}^{\left(12|\ldots\right)\pm},\nn
	C_{21|3\ldots n}^{\pm}= & \sum_{l=0}^{2}\sum_{m=0}^{2-l}c_{l,m,0}^{\left(21|\ldots\right)\pm}\,t^{l}\,y^{m}+\mu^{2}c_{0,0,2}^{\left(21|\ldots\right)\pm},
	\end{align}
	one can verify that the coefficients of $C_{21|3\ldots n}^{\pm}$ are completely determined by
	\begin{align}
	& c_{l,m,0}^{\left(21|\ldots\right)\pm}=\sum_{r=l}^{2}\left(\sum_{s=\max[0,l+m-r]}^{\min[m,2-r]}\alpha_{l-r,m-s,0}^{\pm}\,c_{r,s,0}^{\left(12|\ldots\right)\pm}\right),
	 & c_{0,0,2}^{\left(21|\ldots\right)\pm}=\alpha_{-2,0,2}^{\pm}\,c_{2,0,0}^{\left(12|\ldots\right)\pm}+\alpha_{0,0,0}^{\pm}\,c_{0,0,2}^{\left(12|\ldots\right)\pm}.
	 \label{cf2}
	\end{align}
\section{Examples}
\label{sec:example}
In this last section we verify on some explicit examples the coefficient relations we have previously derived.\\
In order to obtain compact expressions and keep the discussion as simple possible, we consider scalar loop contributions to gluon amplitudes only and we present analytic results for convenient helicity configurations.
Nevertheless, numerical checks of the coefficient relations have been performed for all helicity configurations and gluon loop contributions have been included as well. All results presented in this section have been numerically validated against the ones provided by the C$++$ library \njet~\cite{Badger:2012pg}.\\
In addition, we would like to mention that, besides constituting one of the FDF ingredients needed for the computation of the full amplitude, the scalar contributions presented in this Section can been used as the generators of rational terms in alternative frameworks, such as supersymmetric decomposition \cite{Bern:1993mq,Bern:1994cg}.\\

\subsection{Pentagons}
\begin{figure}[h]
	\centering
	\vspace*{0.5cm}
	\includegraphics[scale=1.1]{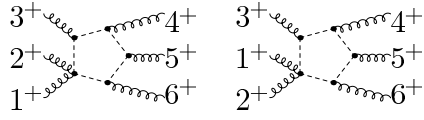}
\caption{Pentagon topologies for the cuts $C_{12|3|4|5|6}$ and $C_{21|3|4|5|6}$.}
\label{figs/5gP.21}
\end{figure}
To begin with, we consider the six-gluon helicity amplitude $\mathcal{A}_{6}^{1\text{-loop}}\left(1^{+},2^{+},3^{+},4^{+},5^{+},6^{+}\right)$ and we compute the quintuple cuts $C^{\pm}_{12|3|4|5|6}$ and $C^{\pm}_{21|3|4|5|6}$ of Fig.~\ref{figs/5gP.21}, whose solutions are parametrized as
\begin{align}
& l_{5}^{+}=\frac{c}{2}\left\langle 4\left|\gamma^{\mu}\right|5\right]-\frac{\mu^{2}}{2s_{45}c}\left\langle 5\left|\gamma^{\mu}\right|4\right], &  & l_{5}^{-}=\frac{c}{2}\left\langle 5\left|\gamma^{\mu}\right|4\right]-\frac{\mu^{2}}{2s_{45}c}\left\langle 4\left|\gamma^{\mu}\right|5\right],
\label{eq:5cCS}
\end{align}
where the parameter $c$ is fixed by the on-shell conditions.
From the products of trees we obtain
\begin{align}
C^{\pm}_{12|3|4|5} & =A_{4}^{\text{tree}}\left(-l^{\pm}_{1},1^{+},2^{+},l^{\pm}_{3}\right)A_{3}^{\text{tree}}\left(-l^{\pm}_{3},3^{+},l^{\pm}_{4}\right)A_{3}^{\text{tree}}\left(-l^{\pm}_{4},4^{+},l^{\pm}_{5}\right)A_{3}^{\text{tree}}\left(-l^{\pm}_{5},5^{+},l^{\pm}_{6}\right)A_{3}^{\text{tree}}\left(-l^{\pm}_{6},6^{+},l^{\pm}_{1}\right)\nonumber \\ &=\frac{i\mu^{2}[2|1]\langle3|l_{5}|4]\langle4|l_{4}|3]\langle5|l_{1}|6]\langle6|l_{6}|5]}{\langle1|2\rangle\langle3|4\rangle^{2}\langle5|6\rangle^{2}\langle1|l_{1}|1]}\label{eq:ex5c12}
\end{align}
and
\begin{align}
C_{21|3|4|5} & =A_{4}^{\text{tree}}\left(-l^{\pm}_{1},2^{+},1^{+},l^{\pm}_{3}\right)A_{3}^{\text{tree}}\left(-l^{\pm}_{3},3^{+},l^{\pm}_{4}\right)A_{3}^{\text{tree}}\left(-l^{\pm}_{4},4^{+},l^{\pm}_{5}\right)A_{3}^{\text{tree}}\left(-l^{\pm}_{5},5^{+},l^{\pm}_{6}\right)A_{3}^{\text{tree}}\left(-l^{\pm}_{6},6^{+},l^{\pm}_{1}\right),\nonumber \\
& =\frac{i\mu^{2}[2|1]\langle3|l_{5}|4]\langle4|l_{4}|3]\langle5|l_{1}|6]\langle6|l_{6}|5]}{\langle1|2\rangle\langle3|4\rangle^{2}\langle5|6\rangle^{2}\langle2|l_{1}|2]}.
\label{eq:ex5c21}
\end{align}
The two cuts are related by the BCJ identity \eqref{eq:pentaC21},
\begin{align}
C^{\pm}_{21|3|4|5} & =\frac{\left(l^{\pm}_{3}+p_{2}\right)^{2}-\mu^{2}}{\left(l^{\pm}_{1}-p_{2}\right)^{2}-\mu^{2}}C{\pm}_{12|3|4|5|6}.
\label{id5pt}
\end{align}
Using  momentum conservation to express $l_{5}^{\pm}$ in terms of $l_{1}^{\pm},l_{3}^{\pm},l_{4}^{\pm}$,
\begin{align}
& l_{1}^{\pm}=l_{5}^{\pm}-p_{5}-p_{6}, &  & l_{3}^{\pm}=l_{5}^{\pm}+p_{3}+p_{4}, &  & l_{4}^{\pm}=l_{5}^{\pm}+p_{4}, &  & l_{6}^{\pm}=l_{5}^{\pm}-p_{5},
\end{align}
one can verify that $C^{\pm}_{12|3|4|5|6}$ takes the form
\begin{align}
C_{12|3|4|5|6}^{\pm} & =\frac{i\mu^{2}s_{34}{}^{2}s_{45}{}^{2}s_{56}{}^{2}[2|1][4|3][6|5]\langle3|1+2|6]^{2}\langle6|1+2|3]^{2}}{\text{tr}_{5}(6,3,5,4)^{3}\langle1|2\rangle\langle3|4\rangle\langle5|6\rangle\left(s_{45}\text{tr}_{5}(1,5,2,6)+s_{345}\text{tr}_{5}(1,5,4,6)-s_{16}\text{tr}_{5}(3,4,5,6)\right)},
\label{anl5pt}
\end{align}
where $s_{ij}=\langle i j\rangle[ji]$ and $\text{tr}_{5}(1,2,3,4)=\left\langle 1\left|234\right|1\right]-\left\langle 1\left|432\right|1\right]$.\\
In a similar way, according to \eqref{ratio5pt}, we find
\begin{align}
\frac{\left(l^{\pm}_{3}+p_{2}\right)^{2}-\mu^{2}}{\left(l^{\pm}_{1}-p_{2}\right)^{2}-\mu^{2}}& =\frac{s_{45}\text{tr}_{5}(1,5,2,6)+s_{345}\text{tr}_{5}(1,5,4,6)-s_{16}\text{tr}_{5}(3,4,5,6)}{s_{45}\text{tr}_{5}(2,5,1,6)+s_{345}\text{tr}_{5}(2,5,4,6)-s_{26}\text{tr}_{5}(3,4,5,6)}.
\label{R5pt}
\end{align}
Hence, substituting \eqref{anl5pt} and \eqref{R5pt} in \eqref{id5pt} we obtain
\begin{align}
C_{21|3|4|5|6}^{\pm} & =\frac{i\mu^{2}s_{34}{}^{2}s_{45}{}^{2}s_{56}{}^{2}[2|1][4|3][6|5]\langle3|1+2|6]^{2}\langle6|1+2|3]^{2}}{\text{tr}_{5}(6,3,5,4)^{3}\langle1|2\rangle\langle3|4\rangle\langle5|6\rangle\left(s_{45}\text{tr5}(2,5,1,6)+s_{345}\text{tr5}(2,5,4,6)-s_{26}\text{tr5}(3,4,5,6)\right)},
\end{align}
which reproduces the same result one could obtain from similar algebraic manipulations on \eqref{eq:ex5c21}. The analytic expressions for the two cuts find numerical agreement with \njet.

\subsection{Boxes}
\begin{figure}[h]
	\centering
	\vspace*{0.5cm}
	\includegraphics[scale=1.1]{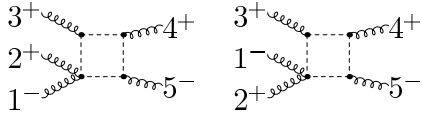}
\caption{Box topologies for the cuts $C_{12|3|4|5}$ and $C_{21|3|4|5}$.}
\label{figs/5gQ.21}
\end{figure}
As an example of identities between box coefficients, we consider the quadruple cuts $C^{\pm}_{12|3|4|5}$
and $C^{\pm}_{21|3|4|5}$ for the helicity amplitude $\mathcal{A}_{5}^{1\text{-loop}}\left(1^{-},2^{+},3^{+},4^{+},5^{-}\right)$, depicted in Fig.~\ref{figs/5gQ.21}.\\
For this configuration the cut solutions can be parametrized as 
\begin{align}
& l_{5}^{+}=\frac{c_{+}}{2}\left\langle 4\left|\gamma^{\mu}\right|5\right]-\frac{\mu^{2}}{2s_{45}c_{+}}\left\langle 5\left|\gamma^{\mu}\right|4\right], &  & l_{5}^{-}=\frac{c_{-}}{2}\left\langle 5\left|\gamma^{\mu}\right|4\right]-\frac{\mu^{2}}{2s_{45}c_{-}}\left\langle 4\left|\gamma^{\mu}\right|5\right],
\end{align}
being $c_{+}$ and $c_{-}$ coefficients determined by behavior of the on-shell solutions at $\mu^2\to 0$ and $\mu^2\to\infty$. \\
By combining tree-level amplitudes we can write
\begin{align}
C^{\pm}_{12|3|4|5} & =A_{4}^{\text{tree}}\left(-l^{\pm}_{1},1^{-},2^{+},l^{\pm}_{3}\right)A_{3}^{\text{tree}}\left(-l^{\pm}_{3},3^{+},l^{\pm}_{4}\right)A_{3}^{\text{tree}}\left(-l^{\pm}_{4},4^{+},l^{\pm}_{5}\right)A_{3}^{\text{tree}}\left(-l^{\pm}_{5},5^{-},l^{\pm}_{1}\right)\nonumber \\
& =\frac{\langle1|l_{1}|2]{}^{2}\langle3|l_{5}|4]\langle4|l_{4}|3]\langle5|l_{1}|1]}{s_{12}[5|1]\langle3|4\rangle^{2}\langle1|l_{1}|1]}
\end{align}
and
\begin{align}
C^{\pm}_{21|3|4|5} & =A_{4}^{\text{tree}}\left(-l^{\pm}_{1},2^{+},1^{-},l^{\pm}_{3}\right)A_{3}^{\text{tree}}\left(-l^{\pm}_{3},3^{+},l^{\pm}_{4}\right)A_{3}^{\text{tree}}\left(-l^{\pm}_{4},4^{+},l^{\pm}_{5}\right)A_{3}^{\text{tree}}\left(-l^{\pm}_{5},5^{-},l^{\pm}_{1}\right)\nonumber \\
& =\frac{\langle1|l_{1}|2]{}^{2}\langle3|l_{5}|4]\langle4|l_{4}|3]\langle5|l_{1}|1]}{s_{12}[5|1]\langle3|4\rangle^{2}\langle2|l_{1}|2]}
\end{align}
the two cuts can be related through \eqref{C21box}
\begin{align}
C^{\pm}_{21|3|4|5} & =\frac{\left(l^{\pm}_{3}+p_{2}\right)^{2}-\mu^{2}}{\left(l^{\pm}_{1}-p_{2}\right)^{2}-\mu^{2}}C^{\pm}_{21|3|4|5}.
\end{align}
Momentum conservation allows us to write,
\begin{align}
& l^{\pm}_{1}=l^{\pm}_{5}-p_{5}, &  & l^{\pm}_{4}=l^{\pm}_{5}+p_{4}, &  & l_{3}=l^{\pm}_{5}+p_{3}+p_{4},
\end{align}
and, consequently, to express $C^{\pm}_{12|3|4|5}$ as
\begin{align}
C_{12|3|4|5}^{\pm} & =-\frac{i\mu^{4}[4|3]\text{tr}_{5}(\eta_{1,2},4,3,5)^{2}}{s_{12}\text{tr}_{5}(3,4,1,5)[5|3][5|4]\langle3|4\rangle^{2}},
\label{id3ex}
\end{align}
where we have introduced the complex momenta $\eta_{i,j}^{\nu}=\frac{1}{2}\left\langle i\left|\gamma^{\nu}\right|j\right]$.\\
We observe that, for this particular helicity configuration, the box coefficient is given by the $\sim \mu^4$ term  only.\\
 Therefore, we just need to compute the ratio of propagators in the large-$\mu^{2}$ limit,
\begin{align}
\left.\frac{\left(l^{\pm}_{3}+p_{2}\right)^{2}-\mu^{2}}{\left(l^{\pm}_{1}-p_{2}\right)^{2}-\mu^{2}}\right|_{\mu^{2}\to\infty} & =-1+\mathcal{O}\left(\frac{1}{\mu}\right).
\end{align}
Thanks to this result, the expression for $C_{21|3|4|5}$ obtained from \eqref{id3ex} is
\begin{align}
C_{21|3|4|5}^{\pm} & =\frac{i\mu^{4}[4|3]\text{tr}_{5}(\eta_{1,2},4,3,5)^{2}}{s_{12}\text{tr}_{5}(3,4,1,5)[5|3][5|4]\langle3|4\rangle^{2}},
\end{align}
which finds again agreement with \njet.

\subsection{Triangles}
\begin{figure}[h]
	\centering
	\vspace*{0.5cm}
	\includegraphics[scale=1.1]{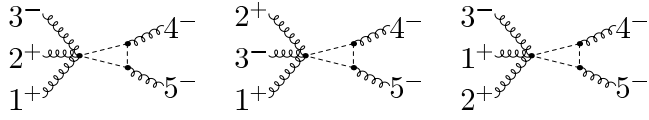}
	\caption{Triangle topologies for the cuts $C_{123|4|5},C_{132|4|5}$ and $C_{213|4|5}$.}
	\label{figs/5gT.213}
\end{figure}
For triple cuts we give an example of coefficient relations obtained through identities between five-point tree-level amplitudes, which are discussed in \ref{apd}.
\\Let us  consider $\mathcal{A}_{5}^{1\text{-loop}}\left(1^{+},2^{+},3^{-},4^{-},5^{-}\right)$  and the three cuts of Fig.~\ref{figs/5gT.213}, $C^{\pm}_{213|4|5}$, $C^{\pm}_{123|4|5}$ and $C^{\pm}_{132|4|5}$, respectively.\\
The cut solutions are given by
\begin{align}
& l_{5}^{+}=\frac{t}{2}\left\langle 4\left|\gamma^{\nu}\right|5\right]-\frac{\mu^{2}}{2s_{45}t}\left\langle 5\left|\gamma^{\nu}\right|4\right], &  & l_{5}^{-}=\frac{t}{2}\left\langle 5\left|\gamma^{\nu}\right|4\right]-\frac{\mu^{2}}{2s_{45}t}\left\langle 4\left|\gamma^{\nu}\right|5\right],
\end{align}
and from the product of trees we obtain
\begin{align}
C^{\pm}_{123|4|5} & =A_{5}^{\text{tree}}\left(-l^{\pm}_{1},1^{+},2^{+},3^{-},l^{\pm}_{4}\right)A_{3}^{\text{tree}}\left(-l^{\pm}_{4},4^{-},l^{\pm}_{5}\right)A_{3}^{\text{tree}}\left(-l^{\pm}_{5},5^{-},l^{\pm}_{1}\right)\nonumber \\
& =\frac{i\langle5|l^{\pm}_{1}|4|l^{\pm}_{5}|5]\langle3|1+2|l^{\pm}_{1}|3\rangle{}^{2}}{s_{123}[5|4]^{2}\langle1|2\rangle\langle2|3\rangle\langle1|l^{\pm}_{1}|1+2|3\rangle}-\frac{i\mu^{2}[2|1]\langle3|l^{\pm}_{4}|2]{}^{2}\langle5|l^{\pm}_{1}|4|l_{5}|5]}{[5|4]^{2}[3|l^{\pm}_{4}|3|2]\langle1|l^{\pm}_{1}|1]\langle1|2+3|l^{\pm}_{4}|3\rangle},\\
C^{\pm}_{132|4|5} & =A_{5}^{\text{tree}}\left(-l^{\pm}_{1},1^{+},3^{-},2^{+},l^{\pm}_{4}\right)A_{3}^{\text{tree}}\left(-l^{\pm}_{4},4^{-},l^{\pm}_{5}\right)A_{3}^{\text{tree}}\left(-l^{\pm}_{5},5^{-},l^{\pm}_{1}\right)\nonumber \\
& =-\frac{i\langle3|l^{\pm}_{1}|1]{}^{2}\langle3|l^{\pm}_{4}|2]{}^{2}\langle5|l^{\pm}_{1}|4|l^{\pm}_{5}|5]}{[5|4]^{2}[2|l^{\pm}_{4}|2+3|1]\langle1|l^{\pm}_{1}|1|3\rangle\langle2|l^{\pm}_{4}|2|3\rangle}-\frac{i\mu^{2}[2|1]^{4}\langle5|l^{\pm}_{1}|4|l^{\pm}_{5}|5]}{s_{123}[3|1][3|2][5|4]^{2}[2|l_{4}|2+3|1]},\\
C^{\pm}_{213|4|5} & =A_{5}^{\text{tree}}\left(-l^{\pm}_{1},2^{+},1^{+},3^{-},l^{\pm}_{4}\right)A_{3}^{\text{tree}}\left(-l^{\pm}_{4},4^{-},l^{\pm}_{5}\right)A_{3}^{\text{tree}}\left(-l^{\pm}_{5},5^{-},l^{\pm}_{1}\right)\nonumber \\
& =-\frac{i\langle5|l^{\pm}_{1}|4|l^{\pm}_{5}|5]\langle3|1+2|l^{\pm}_{1}|3\rangle{}^{2}}{s_{123}[5|4]^{2}\langle1|2\rangle\langle1|3\rangle\langle2|l^{\pm}_{1}|1+2|3\rangle}+\frac{i\mu^{2}[2|1]\langle3|l^{\pm}_{4}|1]{}^{2}\langle5|l^{\pm}_{1}|4|l^{\pm}_{5}|5]}{[5|4]^{2}[3|l^{\pm}_{4}|3|1]\langle2|l^{\pm}_{1}|2]\langle2|1+3|l^{\pm}_{4}|3\rangle}
\end{align}
The three cuts are related by the BCJ identity \eqref{BCJ5pt1},
\begin{align}
C^{\pm}_{213|4|5} & =\frac{\left(P_{l^{\pm}_{4}2}^{2}-\mu^{2}+P_{23}^{2}\right)}{\left(P_{-l^{\pm}_{1}2}^{2}-\mu^{2}\right)}C^{\pm}_{123|4|5}+\frac{\left(P_{l^{\pm}_{4}2}^{2}-\mu^{2}\right)}{\left(P_{-l^{\pm}_{1}2}^{2}-\mu^{2}\right)}C^{\pm}_{132|4|5}\label{eq:ex3c213}.
\end{align}
By using momentum conservation,
\begin{align}
& l_{1}^{\pm}=l_{5}^{\pm}-p_{5}, &  & l_{4}^{\pm}=l_{5}^{\pm}+p_{4},
\end{align}
and expanding $C^{\pm}_{123|4|5}$ and $C^{\pm}_{132|4|5}$ for $t\to\infty$, we obtain
\begin{subequations}
	\begin{align}
	C_{123|4|5}^{+}\left(t,\mu^{2}\right) & =\frac{i\mu^{2}\langle3|4\rangle^{2}(\langle1|4\rangle\langle3|5\rangle+\langle1|3\rangle\langle4|5\rangle)}{[5|4]\langle1|2\rangle\langle1|4\rangle^{2}\langle2|3\rangle}-\frac{i\mu^{2}\langle3|4\rangle^{3}}{[5|4]\langle1|2\rangle\langle1|4\rangle\langle2|3\rangle}t,\\
	C_{123|4|5}^{-}\left(t,\mu^{2}\right) & =\frac{i\mu^{2}\langle3|4\rangle\langle3|5\rangle^{2}}{[5|4]\langle1|2\rangle\langle1|5\rangle\langle2|3\rangle}+\frac{i\mu^{2}\langle3|5\rangle^{3}}{[5|4]\langle1|2\rangle\langle1|5\rangle\langle2|3\rangle}t,\\
	C_{132|4|5}^{+}\left(t,\mu^{2}\right) & =-\frac{i\mu^{2}\langle3|4\rangle^{3}(\langle1|4\rangle\langle3|5\rangle+\langle1|3\rangle\langle4|5\rangle)}{[5|4]\langle1|3\rangle\langle1|4\rangle^{2}\langle2|3\rangle\langle2|4\rangle}+\frac{i\mu^{2}\langle3|4\rangle^{4}}{[5|4]\langle1|3\rangle\langle1|4\rangle\langle2|3\rangle\langle2|4\rangle}t,\\
	C_{132|4|5}^{-}\left(t,\mu^{2}\right) & =-\frac{i\mu^{2}(\langle2|5\rangle\langle3|4\rangle-\langle2|3\rangle\langle4|5\rangle)\langle3|5\rangle^{3}}{[5|4]\langle1|3\rangle\langle1|5\rangle\langle2|3\rangle\langle2|5\rangle^{2}}-\frac{i\mu^{2}\langle3|5\rangle^{4}}{[5|4]\langle1|3\rangle\langle1|5\rangle\langle2|3\rangle\langle2|5\rangle}t.
		\label{eq:ex3cCs}
	\end{align}
	\end{subequations}
In a similar way, the expansion for large-$t$ of the ratio of propagators returns
 \begin{subequations} 
	\begin{align}
	\left.\frac{\left(P_{l^{+}_{4}2}^{2}-\mu^{2}+P_{23}^{2}\right)}{\left(P_{-l^{+}_{1}2}^{2}-\mu^{2}\right)}\right|_{t\to\infty}& =\frac{\mu^{2}s_{12}s_{24}s_{25}}{s_{45}t^{3}\langle4|2|5]^{3}}+\frac{s_{12}s_{25}{}^{2}}{t^{3}\langle4|2|5]^{3}}+\frac{s_{12}s_{25}}{t^{2}\langle4|2|5]^{2}}+\frac{s_{12}}{t\langle4|2|5]}-1+\mathcal{O}\left(\frac{1}{t^{4}}\right),\\
	\left.\frac{\left(P_{l^{-}_{4}2}^{2}-\mu^{2}+P_{23}^{2}\right)}{\left(P_{-l^{-}_{1}2}^{2}-\mu^{2}\right)}\right|_{t\to\infty} & =\frac{\mu^{2}s_{12}s_{24}s_{25}}{s_{45}t^{3}\langle5|2|4]^{3}}+\frac{s_{12}s_{25}{}^{2}}{t^{3}\langle5|2|4]^{3}}+\frac{s_{12}s_{25}}{t^{2}\langle5|2|4]^{2}}+\frac{s_{12}}{t\langle5|2|4]}-1+\mathcal{O}\left(\frac{1}{t^{4}}\right),\\
	\left.\frac{\left(P_{l^{+}_{4}2}^{2}-\mu^{2}\right)}{\left(P_{-l^{+}_{1}2}^{2}-\mu^{2}\right)}\right|_{t\to\infty}& =-\frac{\mu^{2}s_{24}s_{25}\left(s_{24}+s_{25}\right)}{s_{45}t^{3}\langle4|2|5]^{3}}-\frac{s_{25}{}^{2}\left(s_{24}+s_{25}\right)}{t^{3}\langle4|2|5]^{3}}-\frac{s_{25}\left(s_{24}+s_{25}\right)}{t^{2}\langle4|2|5]^{2}}-\frac{s_{24}+s_{25}}{t\langle4|2|5]}-1+\mathcal{O}\left(\frac{1}{t^{4}}\right),\\
	\left.\frac{\left(P_{l^{-}_{4}2}^{2}-\mu^{2}\right)}{\left(P_{-l^{-}_{1}2}^{2}-\mu^{2}\right)}\right|_{t\to\infty} & =-\frac{\mu^{2}s_{24}s_{25}\left(s_{24}+s_{25}\right)}{s_{45}t^{3}\langle5|2|4]^{3}}-\frac{s_{25}{}^{2}\left(s_{24}+s_{25}\right)}{t^{3}\langle5|2|4]^{3}}-\frac{s_{25}\left(s_{24}+s_{25}\right)}{t^{2}\langle5|2|4]^{2}}-\frac{s_{24}+s_{25}}{t\langle5|2|4]}-1+\mathcal{O}\left(\frac{1}{t^{4}}\right).
	\end{align}
	\label{eq:ex3cRs}\end{subequations}
Thus, plugging these results into \eqref{eq:ex3c213} we can obtain
\begin{align}
C_{213|4|5}^{+}\left(t,\mu^{2}\right) & =-\frac{i\mu^{2}\langle3|4\rangle^{2}(\langle2|4\rangle\langle3|5\rangle+\langle2|3\rangle\langle4|5\rangle)}{[5|4]\langle1|2\rangle\langle1|3\rangle\langle2|4\rangle^{2}}+\frac{i\mu^{2}\langle3|4\rangle^{3}}{[5|4]\langle1|2\rangle\langle1|3\rangle\langle2|4\rangle}t,\\
C_{213|4|5}^{-}\left(t,\mu^{2}\right) & =-\frac{i\mu^{2}\langle3|4\rangle\langle3|5\rangle^{2}}{[5|4]\langle1|2\rangle\langle1|3\rangle\langle2|5\rangle}-\frac{i\mu^{2}\langle3|5\rangle^{3}}{[5|4]\langle1|2\rangle\langle1|3\rangle\langle2|5\rangle}t,
\end{align}
which agrees with the $t\to\infty$ expansion of \eqref{eq:ex3cCs}. The contributions of the three cuts to $\mathcal{A}_{5}^{1\text{-loop}}\left(1^{+},2^{+},3^{-},4^{-},5^{-}\right)$  have been numerically checked with \njet.


\subsection{Bubbles}
\begin{figure}[h]
	\centering
	\vspace*{0.5cm}
	\includegraphics[scale=1.1]{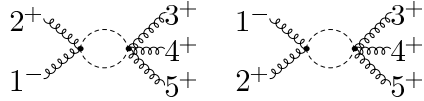}
\caption{Bubble topologies for the cuts $C_{12|345}$ and $C_{21|345}$.}
\label{figs/5gD.21.tex}
\end{figure}
As a final example, we compute the double cuts $C_{12|345}$ and $C_{21|345}$
of the helicity amplitude $\mathcal{A}_{5}^{1\text{-loop}}\left(1^{-},2^{+},3^{+},4^{+},5^{+}\right)$, which are depicted in Fig.~\ref{figs/5gD.21.tex}. For sake of simplicity, we will consider only pure bubble contributions but we remark spurious terms originating from triangles, which should subtracted in order to recover the full integral coefficient, can be related through BCJ identities in the same way, as discussed in Section~\ref{subbubble}.\\
The cut solutions are given by
\begin{align}
\left(l_{1}^{\mu}\right)^{+} & =y\,e_{1}^{\nu}+\frac{\left(1-y\right)s_{12}}{s_{14}+s_{24}}\,p_{4}^{\nu}+t\,e_{3}^{\nu}+\frac{\left(1-y\right)ys_{12}-\mu^{2}}{\left(s_{14}+s_{24}\right)t}\,e_{4}^{\nu},\\
\left(l_{1}^{\mu}\right)^{-} & =y\,e_{1}^{\nu}+\frac{\left(1-y\right)s_{12}}{s_{14}+s_{24}}\,p_{4}^{\nu}+t\,e_{4}^{\nu}+\frac{\left(1-y\right)ys_{12}-\mu^{2}}{\left(s_{14}+s_{24}\right)t}\,e_{3}^{\nu},
\end{align}
being 
\begin{align}
& e_{1}^{\nu}=\left(p_{1}+p_{2}\right)^{\nu}-\frac{s_{12}}{s_{14}s_{24}}p_{4}^{\nu}, &  & e_{3}^{\nu}=\frac{1}{2}\left\langle e_{1}\left|\gamma^{\nu}\right|4\right], &  & e_{4}^{\nu}=\frac{1}{2}\left\langle 4\left|\gamma^{\nu}\right|e_{1}\right].
\end{align}
By combining tree amplitudes, we can write the two cuts as
\begin{align}
C^{\pm}_{12|345} & =A_{4}^{\text{tree}}\left(-l^{\pm}_{1},1^{-},2^{+},l^{\pm}_{3}\right)A_{5}^{\text{tree}}\left(-l^{\pm}_{3},3^{+},4^{+},5^{+},l^{\pm}_{1}\right)=\frac{\mu^{2}[5|3+4|l^{\pm}_{3}|3]\langle1|l^{\pm}_{1}|2]{}^{2}}{s_{12}\langle3|4\rangle\langle4|5\rangle\langle1|l^{\pm}_{1}|1]\langle3|l^{\pm}_{3}|3]\langle5|l^{\pm}_{1}|5]},\label{eq:bub1}\\
C^{\pm}_{21|345} & =A_{4}^{\text{tree}}\left(-l^{\pm}_{1},2^{+},1^{-},l^{\pm}_{3}\right)A_{5}^{\text{tree}}\left(-l^{\pm}_{3},3^{+},4^{+},5^{+},l^{\pm}_{1}\right)=\frac{\mu^{2}[5|3+4|l_{3}|3]\langle1|l^{\pm}_{1}|2]{}^{2}}{s_{12}\langle3|4\rangle\langle4|5\rangle\langle2|l^{\pm}_{1}|2]\langle3|l^{\pm}_{3}|3]\langle5|l^{\pm}_{1}|5]}\label{eq:bub2}
\end{align}
and, according to \eqref{eq:bcj4pt2c}, they can be related through 
\begin{align}
C^{\pm}_{21|345}= & \frac{\left(l^{\pm}_{3}+p_{2}\right)^{2}-\mu^{2}}{\left(l^{\pm}_{1}-p_{2}\right)^{2}-\mu^{2}}\,C^{\pm}_{12|345}.
\label{idbuex}
\end{align}
By making use of $l_{3}^{\pm}=l_{1}^{\pm}-p_{1}-p_{2}$ and by expanding \eqref{eq:bub1} in the large-$t$ limit, we get
\begin{align}
C_{12|345}^{+} & =\mu^{2}\frac{i[4|2]^{3}\langle1|2\rangle}{s_{34}s_{45}[4|1]\langle3|5\rangle^{2}},\\
C_{12|345}^{-} & =\mu^{2}\frac{i\langle1|4\rangle^{3}}{\langle1|2\rangle\langle2|4\rangle\langle3|4\rangle^{2}\langle4|5\rangle^{2}},
\end{align}
whereas the expansion of the ratios of propagators reads
\begin{align}
\left.\frac{\left(l^{+}_{3}+p_{2}\right)^{2}-\mu^{2}}{\left(l^{+}_{1}-p_{2}\right)^{2}-\mu^{2}}\right|_{t\to\infty} & =-\frac{\left(s_{14}-s_{24}\right)[2|1]^{2}[4|e_{1}]^{2}}{\left(s_{14}+s_{24}\right)[4|1]^{2}[4|2]^{2}}\frac{y}{t^{2}}-\frac{[2|1]^{2}\langle2|4\rangle[4|e_{1}]^{2}}{\left(s_{14}+s_{24}\right)t^{2}[4|1]^{2}[4|2]}\frac{1}{t^{2}}-\frac{[2|1][4|e_{1}]}{[4|1][4|2]}\frac{1}{t}-1+\mathcal{O}\left(\frac{1}{t^{3}}\right),\\
\left.\frac{\left(l^{-}_{3}+p_{2}\right)^{2}-\mu^{2}}{\left(l^{-}_{1}-p_{2}\right)^{2}-\mu^{2}}\right|_{t\to\infty} & =-\frac{\left(s_{14}-s_{24}\right)\langle1|2\rangle^{2}\langle e_{1}|4\rangle^{2}}{\left(s_{14}+s_{24}\right)\langle1|4\rangle^{2}\langle2|4\rangle^{2}}\frac{y}{t^{2}}-\frac{[4|2]\langle1|2\rangle^{2}\langle e_{1}|4\rangle^{2}}{\left(s_{14}+s_{24}\right)\langle1|4\rangle^{2}\langle2|4\rangle}\frac{1}{t^{2}}-\frac{\langle1|2\rangle\langle e_{1}|4\rangle}{\langle1|4\rangle\langle2|4\rangle}\frac{1}{t}-1+\mathcal{O}\left(\frac{1}{t^{3}}\right).
\end{align}
Thanks to these expansions, we can  obtain the analytic expression of $C^{\pm}_{21|345}$ from \eqref{idbuex},
 \begin{subequations}
	\begin{align}
	C_{21|345}^{+} & =-\mu^{2}\frac{i[4|2]^{3}\langle1|2\rangle}{s_{34}s_{45}[4|1]\langle3|5\rangle^{2}}=-C_{12|345}^{+},\\
	C_{21|345}^{-} & =-\mu^{2}\frac{i\langle1|4\rangle^{3}}{\langle1|2\rangle\langle2|4\rangle\langle3|4\rangle^{2}\langle4|5\rangle^{2}}=-C_{12|345}^{-},
	\end{align}
	\label{eq:ex2c21}\end{subequations} 
which agree with what could have been obtained by applying the large-$t$ expansion to \eqref{eq:bub2}.
\section{Conclusions}
In this paper we have presented a set of relations between the coefficients appearing in the decomposition of one-loop QCD amplitudes in terms of master integrals, which have been derived as a byproduct of the color-kinematics duality satisfied by tree-level amplitudes.\\
The complete decomposition of a general one-loop amplitude can be obtained via the $d$-dimensional integrand reduction algorithm, which can be used to express the amplitude in terms of a known basis of loop integrals, whose coefficient can be extracted through suitable Laurent expansions of the integrand evaluated on the on-shell solutions.\\
Since the on-shell integrand factorizes into a product of tree-level amplitudes, BCJ identities between tree-levels have been exploited in order to establish relations between the integral coefficients themselves.\\
In order to be consistent with $d$-dimensional unitarity, hence to obtain a set of identities valid for both cut-constructible part and rational terms, we have made use of BCJ identities for dimensionally regulated trees, which have been derived working in the Four-Dimensional-Formulation scheme (FDF), where the effects of dimensional regularization are carried by massive degrees of freedom.\\
The coefficients identities derived in this paper have been verified on a number of contributions to multi-gluon scattering amplitudes, for which we have provided analytic expressions of the integral coefficients.\\
A natural extension of this work would be the study of similar relations for higher-point one-loop amplitudes, which would require $d$-dimensional BCJ identities between tree-levels with more than five external particles.\\
Moreover, it would be interesting to investigate higher-loop coefficient relations that are expected to descend from BCJ identities at tree-level. To this end, future work will require, beside a general parametrization of the residues of multi-loop integrands, BCJ identities between FDF amplitudes involving more than two external massive particles.
\section*{Acknowledgements}
We wish to thank Pierpaolo Mastrolia for countless discussions and unwavering support during the completion of this work.
W.J.T. also acknowledges Raffaele Fazio for useful discussions, and the University of Edinburgh for kind hospitality while parts of this project were completed.\\
W.J.T. is supported by Fondazione Cassa di Risparmio di Padova e Rovigo (CARIPARO).\\
This work is also partially supported by Padua University Project CPDA144437.\\
The Feynman diagrams depicted in this paper are generated using \Feynarts~\cite{Hahn:2000kx}.
\appendix
\section{Coefficient relations from 5-point BCJ identities}
\label{apd}
In this Appendix we collect the set of identities, obtained through the use of $d$-dimensional BCJ relations for five-point amplitudes of the type \eqref{BCJ5pt1}, that can be used to relate integral coefficients associated to multiple cuts which, beside sharing the same on-shell solutions, differ from the ordering of three external particles.
\subsection{Relations for pentagon coefficients}
\begin{figure}[h]
	\centering
	\vspace*{0.5cm}
	\includegraphics[scale=1.1]{fig10.epsi}
	\caption{Pentagon topologies for the cuts $	C_{123|4\ldots k|\left(k+1\right)\ldots l|\left(l+1\right)\ldots m|\left(m+1\right)\ldots n}$,  $	C_{132|4\ldots k|\left(k+1\right)\ldots l|\left(l+1\right)\ldots m|\left(m+1\right)\ldots n}$ and $C_{231|4\ldots k|\left(k+1\right)\ldots l|\left(l+1\right)\ldots m|\left(m+1\right)\ldots n}$.}
	\label{figs/5sP.123}
\end{figure}
We consider the three quintuple-cuts shown in Fig~\ref{figs/5sP.123}, which differ from the ordering of the particles $p_1$, $p_2$, $p_3$. The contribution from the ordering $\{1,2,3\}$ is given by
\begin{align}
C_{123|4\ldots r|\left(r+1\right)\ldots s|\left(s+1\right)\ldots t|\left(t+1\right)\ldots n}^{\pm}= & A_{5}^{\text{tree}}\left(-l_{1}^{\pm},1,2,3,l_{4}^{\pm}\right)A_{r-1}^{\text{tree}}\left(-l_{4}^{\pm},P_{4\cdots r},l_{r+1}^{\pm}\right)A_{s-r+2}^{\text{tree}}\left(-l_{r+1}^{\pm},P_{r+1\ldots,s},l_{s+1}^{\pm}\right)\nonumber \\
& \times A_{s-t+2}^{\text{tree}}\left(-l_{s+1}^{\pm},P_{s+1\cdots t},l_{t+1}^{\pm}\right)A_{n-t+2}^{\text{tree}}\left(-l_{t+1}^{\pm},P_{t+1\cdots n},l_{1}^{\pm}\right)
\end{align}
and the other two cuts are obtained from the corresponding permutations of $\left\{ 1,2,3\right\} $.\\
Eq.~\eqref{BCJ5pt1} can be used in order to relate the amplitudes $A_{5}^{\text{tree}}\left(-l_{1}^{\pm},1,2,3,l_{4}^{\pm}\right)$, $A_{5}^{\text{tree}}\left(-l_{1}^{\pm},1,3,2,l_{4}^{\pm}\right)$
and $A_{5}^{\text{tree}}\left(-l_{1}^{\pm},2,1,3,l_{4}^{\pm}\right)$ and, thus, to identify
\begin{multline}
C_{213|4\ldots r|\left(r+1\right)\ldots s|\left(s+1\right)\ldots t|\left(t+1\right)\ldots n}^{\pm} 
\\ =\frac{\left(P_{l_{4}^{\pm}2}^{2}+P_{23}^{2}-\mu^{2}\right)C_{123|4\ldots r|\left(r+1\right)\ldots s|\left(s+1\right)\ldots t|\left(t+1\right)\ldots n}^{\pm}+\left(P_{l_{4}^{\pm}2}^{2}-\mu^{2}\right)C_{132|4\ldots r|\left(r+1\right)\ldots s|\left(s+1\right)\ldots t|\left(t+1\right)\ldots n}^{\pm}}{\left(P_{-l_{1}^{\pm}2}^{2}-\mu^{2}\right)}.
\label{eqAp:1}
\end{multline}
Analogously to the case discussed in Section~\ref{subpenta}, the constant ratios of propagators 
\begin{align}
\frac{P_{l_{4}^{\pm}2}^{2}-\mu^{2}}{P_{-l_{1}^{\pm}2}^{2}-\mu^{2}} & =\alpha^{\pm},&
\frac{P_{l_{4}^{\pm}2}^{2}+P_{23}^{2}-\mu^{2}}{P_{-l_{1}^{\pm}2}^{2}-\mu^{2}} & =\beta^{\pm},
\end{align}
allow us to translate \eqref{eqAp:1} into a simple identity between the coefficients of the expansion~\eqref{pentacoef} for the three cuts,
\begin{align}
c^{\left(213|\ldots\right)\pm} & =\beta^{\pm}c^{\left(123|\ldots\right)\pm}+\alpha^{\pm}c^{\left(132|\ldots\right)\pm}.
\end{align}
\subsection{Relations for box coefficients}
\begin{figure}[h]
	\centering
	\vspace*{0.5cm}
	\includegraphics[scale=1.1]{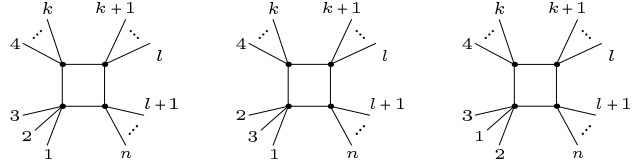}
	\caption{Box topologies for the cuts $C_{123|4\ldots k|k+1\ldots l|l+1\ldots n}$, $C_{132|4\ldots k|k+1\ldots l|l+1\ldots n}$ and $C_{231|4\ldots k|k+1\ldots l|l+1\ldots n}$.}
	\label{figs/5sQ.123}
\end{figure}
Similarly to the previous case, we can use BCJ identities to relate the quadruple cuts depicted in Fig.~\ref{figs/5sQ.123}, given by
\begin{align}
C_{123|4\ldots r|\left(r+1\right)\ldots s|\left(s+1\right)\ldots n}^{\pm}= & A_{5}^{\text{tree}}\left(-l_{1}^{\pm},1,2,3,l_{4}^{\pm}\right)A_{r-1}^{\text{tree}}\left(-l_{4}^{\pm},P_{4\cdots r},l_{r+1}^{\pm}\right)
\nn&\qquad \times A_{s-r+2}^{\text{tree}}\left(-l_{r+1}^{\pm},P_{r+1\ldots,s},l_{s+1}^{\pm}\right)A_{n-s+2}^{\text{tree}}\left(-l_{s+1}^{\pm},P_{s+1\cdots n},l_{1}^{\pm}\right)
\end{align}
and suitable permutations of $\{1,2,3\}$ for $C_{132|4\ldots r|\left(r+1\right)\ldots s|\left(s+1\right)\ldots n}^{\pm}$
and $C_{213|4\ldots r|\left(r+1\right)\ldots s|\left(s+1\right)\ldots n}^{\pm}$.\\
If we make use of~\eqref{BCJ5pt1} for the amplitudes involving the particles $p_1$, $p_2$ and $p_3$, we obtain 
\begin{multline}
C_{213|4\ldots r|\left(r+1\right)\ldots s|\left(s+1\right)\ldots n}^{\pm}  \\
=\frac{\left(P_{l_{4}^{\pm}2}^{2}+P_{23}^{2}-\mu^{2}\right)C_{123|4\ldots r|\left(r+1\right)\ldots s|\left(s+1\right)\ldots n}^{\pm}+\left(P_{l_{4}^{\pm}2}^{2}-\mu^{2}\right)C_{132|4\ldots r|\left(r+1\right)\ldots s|\left(s+1\right)\ldots n}^{\pm}}{\left(P_{-l_{1}^{\pm}2}^{2}-\mu^{2}\right)}.
\label{eqAp:2}
\end{multline}
As shown in Section ~\ref{subbox}, the two box coefficients contributing to the amplitude can be extracted by taking the $\mu^{2}\rightarrow0$ and $\mu^{2}\rightarrow\infty$
limits, where the ratios of propagators behave like
\begin{align}
\left.\frac{P_{l_{4}^{\pm}2}^{2}-\mu^{2}}{P_{-l_{1}^{\pm}2}^{2}-\mu^{2}}\right|_{\mu^{2}\rightarrow0}=\alpha_{0}^{\pm},&&\left.\frac{P_{l_{4}^{\pm}2}^{2}-\mu^{2}}{P_{-l_{1}^{\pm}2}^{2}-\mu^{2}}\right|_{\mu^{2}\rightarrow\infty}=\alpha_{4}^{\pm}+\mathcal{O}\left(\frac{1}{\mu}\right),\nn
\left.\frac{P_{l_{4}^{\pm}2}^{2}+P_{23}^{2}-\mu^{2}}{P_{-l_{1}^{\pm}2}^{2}-\mu^{2}}\right|_{\mu^{2}\rightarrow0}=\beta_{0}^{\pm},&&\left.\frac{P_{l_{4}^{\pm}2}^{2}+P_{23}^{2}-\mu^{2}}{P_{-l_{1}^{\pm}2}^{2}-\mu^{2}}\right|_{\mu^{2}\rightarrow\infty}=\beta_{4}^{\pm}+\mathcal{O}\left(\frac{1}{\mu}\right).
\end{align}
Thus, starting from \eqref{eqAp:2} we can relate the coefficients of the expansions \eqref{exbox1}-\eqref{exbox2} of the three quadruple cuts trough the identities
\begin{align}
c_{i}^{\left(213|\ldots\right)\pm} & =\beta_{i}^{\pm}c_{i}^{\left(123|\ldots\right)\pm}+\alpha_{i}^{\pm}c_{i}^{\left(132|\ldots\right)\pm},\qquad i=0,4.
\end{align}
\subsection{Relations for triangle coefficients}
\begin{figure}[h]
	\centering
	\vspace*{0.5cm}
	\includegraphics[scale=1.1]{fig12.epsi}
	\caption{Triangle topologies for the cuts $C_{123|4\ldots k|\left(k+1\right)\ldots n}$, $C_{132|4\ldots k|\left(k+1\right)\ldots n}$ and $C_{213|4\ldots k|\left(k+1\right)\ldots n}$.}
	\label{figs/5sT.122}
\end{figure}
Now we turn our attention to the triangle topologies shown in Fig.~\ref{figs/5sT.122}. The expression of the cut with external ordering $\{1,2,3,\,...\,n\}$ in terms of tree-level amplitudes is given by
\begin{align}
C^{\pm}_{123|4\ldots k|\left(k+1\right)\ldots n} & =A_{5}^{\text{tree}}\left(-l^{\pm}_{1},1,2,3,l^{\pm}_{4}\right)A_{k-1}^{\text{tree}}\left(-l^{\pm}_{4},P_{4\cdots k},l^{\pm}_{k+1}\right)A_{n-k+2}^{\text{tree}}\left(-l^{\pm}_{k+1},P_{k+1\ldots,n},l^{\pm}_{1}\right)
\end{align}
and, as usual, $C^{\pm}_{132|4\ldots k|\left(k+1\right)\ldots n}$ and $C^{\pm}_{213|4\ldots k|\left(k+1\right)\ldots n}$ are obtained from the corresponding permutations of $\{1,2,3\}$.\\

Eq.~\eqref{BCJ5pt1} allow us to identify 
\begin{align}
C^{\pm}_{213|4\ldots k|\left(k+1\right)\ldots n} & =\frac{\left(P_{l^{\pm}_{4}2}^{2}+P_{23}^{2}-\mu^{2}\right)C^{\pm}_{123|4\ldots k|\left(k+1\right)\ldots n}+\left(P_{l^{\pm}_{4}2}^{2}-\mu^{2}\right)C^{\pm}_{132|4\ldots k|\left(k+1\right)\ldots n}}{\left(P_{-l^{\pm}_{1}2}^{2}-\mu^{2}\right)}
\label{5ptidtri}
\end{align}
and, following the procedure of Section~\ref{subtri}, we can take the large-$t$ limit of the two ratios of propagators,
\begin{align}
\left.\frac{P_{l^{\pm}_{4}2}^{2}-\mu^{2}}{P_{-l^{\pm}_{1}2}^{2}-\mu^{2}}\right|_{t\to\infty} & =\sum_{m=-3}^{0}\alpha_{m,0}^{\pm}t^{m}+\mu^{2}\sum_{m=-3}^{-2}\alpha_{m,2}^{\pm}t^{m}+\mathcal{O}\left(\frac{1}{t^{4}}\right),\nn
\left.\frac{P_{l^{\pm}_{4}2}^{2}+P_{23}^{2}-\mu^{2}}{P_{-l^{\pm}_{1}2}^{2}-\mu^{2}}\right|_{t\to\infty} & =\sum_{m=-3}^{0}\beta_{m,0}^{\pm}t^{m}+\mu^{2}\sum_{m=-3}^{-2}\beta_{m,2}^{\pm}t^{m}+\mathcal{O}\left(\frac{1}{t^{4}}\right),
\end{align}
and use it in \eqref{5ptidtri} in order to express the coefficients of the expansion \eqref{eq:3cute} of $C_{213|4\ldots k|\left(k+1\right)\ldots n}^{\pm}$ in terms of the ones of $C_{123|4\ldots k|\left(k+1\right)\ldots n}$ and $C_{132|4\ldots k|\left(k+1\right)\ldots n}$,
\begin{align}
c_{m,0}^{\left(213|\ldots\right)\pm}= & \sum_{l=0}^{3-m}\left[\beta_{-l,0}^{\pm}\,c_{l+m,0}^{\left(123|\ldots\right)\pm}+\alpha_{-l,0}^{\pm}\,c_{l+m,0}^{\left(132|\ldots\right)\pm}\right],\\
c_{m,2}^{\left(213|\ldots\right)\pm}= & \sum_{l=0}^{1-m}\left[\beta_{-l-2,2}^{\pm}\,c_{l+m+2,0}^{\left(123|\ldots\right)\pm}+\beta_{-l,0}^{\pm}\,c_{l+m,2}^{\left(123|\ldots\right)\pm}+\alpha_{-l-2,2}^{\pm}\,c_{l+m+2,0}^{\left(132|\ldots\right)\pm}+\alpha_{-l,0}^{\pm}\,c_{l+m,2}^{\left(132|\ldots\right)\pm}\right].
\label{cf3}
\end{align}
\subsection{Relations for bubble coefficients}
\begin{figure}[h]
	\centering
	\vspace*{0.5cm}
	\includegraphics[scale=1.1]{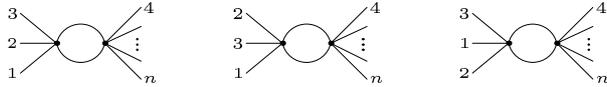}
\caption{Bubble topologies for the cuts $C_{123|4\ldots n}$, $C_{132|4\ldots n}$ and $C_{213|4\ldots n}$.}
\label{figs/5sD.123.tex}
\end{figure}
Finally, we use BCJ identities in order to determine relations between the coefficients of the bubble contributions shown in Fig.~\ref{figs/5sD.123.tex}. The double cut with external ordering $\{1,2,3\,...\,,n\}$ is given by
\begin{align}
C_{123|4\ldots n}^{\pm}=&A_{5}^{\text{tree}}\left(-l_{1}^{\pm},1,2,3,l_{4}^{\pm}\right)A_{n-1}^{\text{tree}}\left(-l_{4}^{\pm},P_{4\cdots n},l_{1}^{\pm}\right),
\end{align}
whereas $C_{132|4\ldots n}^{\pm}$ and $C_{213|4\ldots n}^{\pm}$ are obtained from the corresponding permutations of $\left\{ 1,2,3\right\} $.\\
Hence, thanks to~\eqref{BCJ5pt1}, we can identify
\begin{align}
C_{213|4\ldots n}^{\pm} & =\frac{\left(P_{l_{4}^{\pm}2}^{2}+P_{23}^{2}-\mu^{2}\right)C_{123|4\ldots n}^{\pm}+\left(P_{l_{4}^{\pm}2}^{2}-\mu^{2}\right)C_{132|4\ldots n}^{\pm}}{\left(P_{-l_{1}^{\pm}2}^{2}-\mu^{2}\right)}.
\label{eqAp:3}
\end{align}
As we did in Section~\ref{subbubble}, after taking the $t\rightarrow\infty$ limit of the two ratios of propagators,
\begin{align}
\left.\frac{P_{l_{4}^{\pm}2}^{2}-\mu^{2}}{P_{-l_{1}^{\pm}2}^{2}-\mu^{2}}\right|_{t\rightarrow\infty}=\sum_{l=-2}^{0}\sum_{m=0}^{-l}\alpha_{l,m,0}^{\pm}\, t^{l}\, y^{m}+\frac{\mu^{2}}{t^{2}}\alpha_{-2,0,2}^{\pm}+\mathcal{O}\left(\frac{1}{t^{3}}\right),\nonumber \\
\left.\frac{P_{l_{4}^{\pm}2}^{2}+P_{23}^{2}-\mu^{2}}{P_{-l_{1}^{\pm}2}^{2}-\mu^{2}}\right|_{t\rightarrow\infty}=\sum_{l=-2}^{0}\sum_{m=0}^{-l}\beta_{l,m,0}^{\pm}\, t^{l}\, y^{m}+\frac{\mu^{2}}{t^{2}}\beta_{-2,0,2}^{\pm}+\mathcal{O}\left(\frac{1}{t^{3}}\right),
\end{align}
we can substitute the expansion \eqref{eq:2cut} for the three cuts in \eqref{eqAp:3} and determine the coefficients of $C_{213|4\ldots n}^{\pm}$ from the knowledge of the ones of  $C_{123|4\ldots n}^{\pm}$
and $C_{132|4\ldots n}^{\pm}$,
\begin{subequations}
	\begin{align}
	c_{l,m,0}^{\left(213|\ldots\right)\pm}= & \sum_{r=l}^{2}\left(\sum_{s=\max[0,l+m-r]}^{\min[m,2-r]}\left(\alpha_{l-r,m-s,0}^{\pm}\, c_{r,s,0}^{\left(132|\ldots\right)\pm}+\beta_{l-r,m-s,0}^{\pm}\, c_{r,s,0}^{\left(123|\ldots\right)\pm}\right)\right),\\
	c_{0,0,2}^{\left(213|\ldots\right)\pm}= & \alpha_{-2,0,2}^{\pm}\, c_{2,0,0}^{\left(132|\ldots\right)\pm}+\alpha_{0,0,0}^{\pm}\, c_{0,0,2}^{\left(132|\ldots\right)\pm}+\beta_{-2,0,2}^{\pm}\, c_{2,0,0}^{\left(123|\ldots\right)\pm}+\beta_{0,0,0}^{\pm}\, c_{0,0,2}^{\left(123|\ldots\right)\pm}.
	\end{align}
\end{subequations}

\bibliographystyle{elsarticle-num}
\bibliography{references}

\end{document}